\documentclass[iopart,rmp,amsmath,amssymb,twocolumn]{revtex4-2}

\usepackage[utf8]{inputenc}
\usepackage{graphicx} 
\usepackage[T1]{fontenc}
\usepackage{subfigure}
\usepackage{xcolor} 
\definecolor{customdarkblue}{rgb}{0.0, 0.0, 0.75}  
\usepackage{dsfont}
\usepackage{bm}
\usepackage{qcircuit}
\usepackage{empheq}
\usepackage{orcidlink}
\usepackage{hyperref}
\hypersetup{
    colorlinks=true,    
    urlcolor=blue,      
    linkcolor=blue,    
    citecolor=customdarkblue     
}
\usepackage{orcidlink}

\newcommand{\ket}[1]{\left\vert#1\right\rangle}
\newcommand{\bra}[1]{\left\langle#1\right\vert}

\newcommand{\ketbra}[2]{\ensuremath{| #1 \rangle\!\langle #2 |}}
\newcommand{\floor}[1]{\lfloor #1 \rfloor}
\renewcommand{\selectlanguage}[1]{}

\setcitestyle{square}
\date{\today}

\begin{document}

\title{Symmetric quantum states: a review of recent progress}

\author{Carlo Marconi$^1$,\orcidlink{0000-0002-0578-3636
}, Guillem Müller-Rigat$^2$,\orcidlink{0000-0003-0589-7956}, Jordi Romero-Pallejà$^3$,\orcidlink{0009-0007-2287-8446}, Jordi Tura$^{4,5}$,\orcidlink{0000-0002-6123-1422}, Anna Sanpera$^{3,6}$,\orcidlink{0000-0002-8970-6127}}
\affiliation{$^1$ Istituto Nazionale di Ottica - Consiglio Nazionale delle Ricerche (INO-CNR), \mbox{Largo Enrico Fermi 6, 50125 Firenze, Italy. \textcolor{blue}{carlo.marconi@ino.cnr.it}} }
\affiliation{$^2$ ICFO-Institut de Ciències \mbox{Fotòniques, The Barcelona Institute of Science and}\\ Technology, Castelldefels (Barcelona) \mbox{08860, Spain.  \textcolor{blue}{guillem.muller@icfo.eu}}}
\affiliation{ $^3$ F\'isica Te\`{o}rica: Informaci\'o i Fen\`{o}mens Qu\`{a}ntics. Departament de F\'isica, Universitat Aut\`{o}noma de \mbox{Barcelona, 08193 Bellaterra, Spain }}
\affiliation{ $^4$ \mbox{Instituut-Lorentz, Universiteit Leiden, P.O. Box 9506, 2300 RA Leiden, The Netherlands }}
\affiliation{$^5$ $\langle aQa^L\rangle$ \mbox{Applied Quantum Algorithms, Universiteit Leiden, The Netherlands}}
\affiliation{$^6$ ICREA, \mbox{Pg. Lluís Companys 23, 08010 Barcelona, Spain}}

\begin{abstract}
Symmetric quantum states are fascinating objects. They correspond to multipartite systems that remain invariant under particle permutations. This symmetry is reflected in their compact mathematical characterisation but also in their unique physical properties: they exhibit genuine multipartite entanglement and notable robustness against noise and perturbations. These features make such states particularly well-suited for a wide range of quantum information tasks. Here, we provide a pedagogic analysis of the mathematical structure and relevant physical properties of this class of states. 
Beyond the theoretical framework, robust tools for certifying and verifying the properties of symmetric states in experimental settings are essential. In this regard, we explore how standard techniques -- such as quantum state tomography, Bell tests, and entanglement witnesses -- can be specifically adapted for symmetric systems.
Next, we provide an up-to-date overview of the most relevant applications in which these states outperform other classes of states in specific tasks. Specifically, we address their central role in quantum metrology, highlight their use in quantum error correction codes, and examine their contribution in computation and communication tasks. Finally, we present the current state-of-the-art in their experimental generation, ranging from systems of cold atoms to implementations via quantum algorithms. We also review the most significant results obtained in the different experimental realizations.
Despite the notable progress made in recent years with regard to the characterisation and application of symmetric quantum states, several intriguing questions remain unsolved. We conclude this review by discussing some of these open problems and outlining promising directions for future research.
\end{abstract}

\keywords{symmetry, permutational invariant systems, quantum correlations}

\maketitle

\tableofcontents

\section{Introduction and scope}
\label{sec:Introduction}

\noindent 
All quantum particles that share the same intrinsic properties must be regarded as \textit{identical}. It is essential, however, to distinguish this notion from that of \textit{indistinguishable} particles. Indeed, if two identical particles are separated by a sufficiently large distance, they may, in practice, be distinguished by their spatial location. 
This observation already suggests that the (in)distinguishability of a set of identical particles is fundamentally connected to their spatial degrees of freedom. When identical particles are brought sufficiently close together such that their quantum wavefunctions overlap, they become indistinguishable, in the sense that no physical measurement can consistently attribute one particle to one particular quantum state. 

In order to account for empirical observations, appropriate constraints must be imposed on the quantum state of a system of indistinguishable particles. This requirement stems from the so-called symmetrisation postulate, which asserts that, in three-dimensional space, the state of a collection of indistinguishable particles must be either completely antisymmetric -- in which case the particles are classified as \textit{fermions} -- or completely symmetric under the exchange of any two subsystems -- in which case the particles are known as \textit{bosons}. Consequently, the quantum states of bosonic systems are often referred to as \textit{symmetric}, and their properties will constitute the principal focus of the present review. 
At this stage, a clarification is in order. 
When dealing with such systems, a typical description of their states can be given using a bosonic algebra, that is, a set of operators $\{a_{i},a_{i}^{\dagger}\}$, where the subscript $i$ refers to a certain degree of freedom (e.g., mode, particle, etc.) such that their commutator satisfies $[a_{i},a_{j}^{\dagger}]=\delta_{ij}$. This description is particularly convenient, for instance, in the context of quantum optics, where the Hilbert space $\mathcal{S}_{N}^{(d)}$ of $N$ photons (bosons) in $d$ modes is the subspace of the full Fock space $\mathcal{H}= (\mathbb{C}^{d})^{\otimes N}$ spanned by the states
\begin{equation}
\label{fock_state}
    \ket{n_{1} \dots n_{d}} = \frac{(a^{n_{1}}_{1})^{\dagger} \dots (a^{n_{d}}_{d})^{\dagger}}{\sqrt{n_{1}! \dots n_{d}!}} \ket{\mbox{vac}}~,
\end{equation}
where $n_{k}$ is the number of photons in mode $k$, $N=\sum_{k=1}^{d}n_{k}$ and $\ket{\mbox{vac}}$ is the vacuum state. However, there exists also an equivalent description where $\mathcal{S}_{N}^{(d)}$ can be identified with the subspace of $\mathcal{H}= (\mathbb{C}^{d})^{\otimes N}$ which is invariant under permutations. 
In this way, the state in Eq.~(\ref{fock_state}) can be equivalently described as
\begin{equation}
    \ket{n_{1} \dots n_{d}} = \sum_{\pi \in \mathcal{G}_{N}}\sqrt{\frac{n_{1}! \dots n_{d}!}{N!}}P_{\pi} (\ket{1}^{\otimes n_{1}} \dots \ket{d}^{\otimes n_{d}})~,
\end{equation}
where the states $\ket{i}\in \mathbb{C}^{d}$ refer to different photonic modes and the sum runs over all the elements $\pi$ of the
permutation group $\mathcal{G}_{N}$ of $N$ objects. In spite of being two admissible representations, the former emphasises the role of bosonic operators while the latter focuses on the particles themselves. Moving beyond the paradigm of quantum optics, in this review, when dealing with symmetric states, we will mostly refer to the latter representation, where these states span the symmetric subspace of some given Hilbert space. 

The significance of symmetric states manifests across a broad spectrum of physical phenomena. One of the earliest and most prominent examples is that of the \textit{Dicke states}, introduced by R.H. Dicke in 1954, which describe collective excitations of $N$ two-level atoms interacting with a single-mode electromagnetic field \cite{dicke_coherence_1954}. These states are central to the Dicke model, which exhibits a superradiant phase transition, a paradigmatic example of collective behaviour of atoms. 
Similarly, the Lipkin-Meshkov-Glick (LMG) model \cite{LMG1,LMG2,LMG3}, originally formulated in the context of nuclear physics, features Dicke states as ground states in its isotropic version, and has been fundamental in studying quantum phase transitions and collective phenomena in many-body systems. 
More recently, Dicke states have gained considerable attention in quantum information science. Their relevance spans a variety of applications, from quantum metrology — where they enable precision measurements beyond the standard quantum limit — to quantum communication and computation protocols, including quantum error correction, quantum secret sharing, and distributed quantum computing, among others.

The aim of this review is to provide a comprehensive overview of the current state of the art concerning the properties and applications of symmetric states in quantum protocols and tasks. While this review also addresses the practical challenges associated with the experimental realization of symmetric states, our primary focus is to address the current state of the art in their theoretical characterization. Often we will refer also to the broader set of permutationally invariant states, which share some properties with symmetric states. The references included herein are intended to be representative of the extensive literature on symmetric quantum states over the past decades, with an emphasis on recent progress.

The manuscript is structured as follows. In Section~\ref{sec:Symmetry}, we introduce the mathematical framework necessary to analyse symmetric quantum systems. Section~\ref{sec:Correlations} delves into the landscape of quantum correlations, encompassing entanglement, nonlocality, and related phenomena. Section~\ref{sec:Characterization} discusses experimental techniques for the certification and detection of symmetric states, while Section~\ref{sec:Applications} reviews their principal applications in quantum information protocols. Next, Section~\ref{sec:Implementations} surveys the experimental platforms and methods employed to generate symmetric states in the laboratory. Finally, we conclude in Section~\ref{sec:Conluding Remarks} by highlighting open questions and unresolved problems that continue to stimulate research in this area.


\section{Indistinguishable particles: symmetry under permutations}
\label{sec:Symmetry}

\noindent Symmetries capture the invariance of certain physical systems under specific transformations. One simple example is the case of rotational symmetry, where the properties of a system are preserved under spatial rotations. More generally, for each symmetry transformation, there exists an inverse operation that cancels its effect and restores the system to its original state. There is also an identity operation that leaves the system unchanged. The composition of any two such transformations yields another symmetry transformation, and this composition is associative. These properties define a \textit{group} $(G, \circ)$, i.e., a set $G=\{g_i\}$  equipped with an operation $\circ$ such that, for any two elements $g_i, g_j$ of $G$, the composition $g_i\circ g_j$ also belongs to $G$. 

Group theory provides the mathematical framework to describe symmetries. The action of a symmetry group on a physical system is characterised by a representation, which assigns to each group element $g_i\in G$ a matrix $M(g_i)$ acting on a vector space $V$. This assignment preserves the group composition law, meaning that $M(g_i \circ g_k)=M(g_i)\cdot M(g_k)$, where the right-hand side denotes matrix multiplication. In this way, an isomorphism can be established between an abstract element of the group $g_{i} \in G$ and their matrix representations $M(g_{i})$, the latter set forming a group with matrix multiplication. In an abuse of notation, we will refer to the vector space $V$ both as the state space of the physical system and as the representation of the group. It is important to emphasise that representations are not unique, an observation that will become relevant in the remainder of this section.

When dealing with bosonic indistinguishable particles, the quantum state of the system must remain invariant, up to a global phase, under any permutation of its constituents. More formally, consider the Hilbert space of $N$ qudits, given by $\mathcal{H} = (\mathbb{C}^{d})^{\otimes N}$ and let $\mathcal{G}_N$ be the symmetric group of $N$ elements. Such group contains $N!$ elements, corresponding to the possible permutations of $N$ objects. A representation of $\mathcal{G}_N$ assigns to each permutation $\pi \in \mathcal{G}_{N}$ a corresponding permutation matrix $P_{\pi}$, which acts on the basis states of $\mathcal{H}$ as follows:
\begin{equation*}
    P_{\pi} \ket{i_{0}} \ket{i_{1}} \cdots \ket{i_{N-1}} = \ket{i_{\pi^{-1}(0)}} \ket{i_{\pi^{-1}(1)}} \cdots \ket{i_{\pi^{-1}(N-1)}}~,
\end{equation*}
where $\ket{i_{k}} \in \mathbb{C}^{d}$ for every $k = 0, \dots, N-1$.

A quantum state $\rho$ acting on $\mathcal{H}$ is said \textit{permutationally invariant} (PI) if, for every $\pi \in \mathcal{G}_{N}$, it satisfies the relation
\begin{equation}
\label{PI}
      \rho = P_{\pi} \rho P_{\pi}^{\dagger}~.
\end{equation}

As previously mentioned, the representation of a group is not unique. Nevertheless, for \textit{finite} groups, that is, those that contain a finite number of elements, it is possible to decompose the vector space $V = (\mathbb{C}^d)^{\otimes N}$ into a direct sum of some special representations called \textit{irreducible}. Irreducible representations are those that admit only two trivial invariant subspaces under the group action: the trivial subspace $\{0\}$ and the entire space $V$ itself. In other words, for every group element $g \in G$ and every vector $v \in V$, the action $M(g) v\in V$ preserves the structure of the space. Remarkably, no other non-trivial subspace remains invariant under the group action. As such, irreducible representations can be thought, intuitively, as those that describe the action of the group in its most elementary form. 

A remarkable result in representation theory is the \textit{Schur-Weyl duality}, which allows to decompose the Hilbert space $\mathcal{H} = (\mathbb{C}^{d})^{\otimes N}$ as
\begin{equation}
\label{schur}
    (\mathbb{C}^{d})^{\otimes N} \cong \bigoplus_{\bm{\lambda}~ \vdash (d,N)} \mathcal{H}_{\bm{\lambda}} \otimes \mathcal{K}_{\bm{\lambda}}~,
\end{equation}
where $\mathcal{H}_{\bm{\lambda}}$ and $\mathcal{K}_{\bm{\lambda}}$ are irreducible representations of $\mathcal{G}_{N}$ and the group $\mathcal{U}_{d}$ of unitary matrices of order $d$, respectively. 
The sum runs over all the partitions $\bm{\lambda}$ of $N$ with at most $d$ elements, that is, $\bm{\lambda} = ( \lambda_{0}, \dots \lambda_{d-1}),~ \lambda_{i} \geq 0~ \forall i~, \sum_{i=0}^{d-1} \lambda_{i} = N$. This result is particularly relevant for PI states, since it allows to represent them in a block-diagonal form in the basis of Eq.(\ref{schur}), where the size of each block is greatly reduced as compared to the $d^{N} \times d^{N}$ representation in the computational basis.  In the case of qubits, i.e., $d=2$, Eq.(\ref{schur}) can be recast more explicitly as
\begin{equation}
\label{dec}
(\mathbb{C}^{2})^{\otimes N} \cong \bigoplus_{J=J_{\emph{min}}}^{N/2} \mathcal{H}_{J} \otimes \mathcal{K}_{J}~,
\end{equation}
where $\dim \mathcal{H}_{J} = 2J+1$, and the dimension of the multiplicity spaces $\mathcal{K}_{J}$ is either $1$, if $J=N/2$, or
\begin{equation}
    \dim \mathcal{K}_{J} = \binom{N}{N/2 - J} - \binom{N}{N/2 - J-1}~.
\end{equation}
As a consequence, any PI state can be decomposed in this basis as
\begin{equation}
\rho = \bigoplus_{J=J_{\emph{min}}}^{N/2} \frac{p_{J}}{\dim \mathcal{K}_{J}} \mathds{1}_{J} \otimes \rho_{J}~, 
\label{eq:PIdecom}
\end{equation}
where the state $\rho_{J}$ and the coefficients $p_{J}$ form a probability distribution. 

Of particular relevance to our analysis is the block corresponding to $J=N/2$, which is spanned by the so-called \textit{Dicke states}. Such states are typically introduced in quantum optics to describe the interaction between a single-mode photon and a system of $N$ spin-$\frac{1}{2}$ particles \cite{dicke_coherence_1954}, where they correspond to the simultaneous eigenvectors of the collective spin operators $J_{z}$ and $\mathbf{J}^{2} = J_x^2+ J_y ^2 + J_z ^2$. Note that the total spin $\mathbf{J}^2$ is invariant under $SU(2)$ transformations $U(\theta) = \exp{[-i\theta\mathbf{v}\cdot \mathbf{J}]}$, revealing an intriguing connection between permutation invariance and the rotation group. Each PI block $J$ in Eq.~\eqref{eq:PIdecom} has well defined total spin $\mathbf{J}^2= J(J+1)$, with the fully symmetric sector corresponding to the maximal value $J= N/2$. Within the symmetric block, $N$-qubit Dicke states are distinguished by the eigenvalues of $J_z$, $M\in \{-N/2,-N/2+1,...,N/2 \}$. Equivalently, $N$-qubit Dicke states can be defined as balanced superpositions of $N$-qubit states with $k = N/2-M$ excitations, i.e., 
\begin{equation}
\label{dicke_qubit}
    \ket{D^{N}_{k}} =  \binom{N}{k}^{-1/2} \sum_{\pi \in \mathcal{G}_{N}} \pi(\ket{0}^{\otimes (N-k)} \ket{1}^{\otimes k})~,
\end{equation}
where the sum runs over all the permutations of $N$ parties. 
Eq.~(\ref{dicke_qubit}) can be generalised to the case of $N$ qudits converting the index $k$ into a vector $\bm{k}=(k_{0}, k_{1}, \dots, k_{d-1})$ which specifies how many qudits occupy the states $\ket{0}, \ket{1}, \dots, \ket{d-1}$, respectively. Hence, $N$-qudit Dicke states take the form
\begin{equation}
    \ket{D^{N}_{\bm{k}}} = C(N,\bm{k})^{-1/2} \sum_{\pi \in \mathcal{G}_{N}}\pi(\ket{0}^{\otimes k_{0}} \ket{1}^{\otimes k_{1}} \cdots  \ket{d-1}^{\otimes k_{d-1}})~,
\end{equation}
where the normalisation factor $C(N,\bm{k})$ is given by
\begin{equation}
    C(N,\bm{k}) = \binom{N}{\bm{k}} = \frac{N!}{k_{0}! k_{1}! \cdots k_{d-1}!}~. 
\end{equation}

\noindent To clarify the notation, the Dicke state $\ket{D^{3}_{(1,1,1)}}$ of $N=3$ qudits with $d=3$ and excitations described by the vector $\bm{k} = (1,1,1)$, takes the form
\begin{equation}
    \ket{D^{3}_{(1,1,1)}} = \frac{1}{\sqrt{6}}(\ket{012} + \ket{021} + \ket{120} + \ket{102} +\ket{201} + \ket{210})~·
\end{equation}

\noindent In the case of two qudits ($N=2$), it is more convenient to represent symmetric states using the following notation for Dicke states:
\begin{align}
\label{two_dicke1}
    \ket{D_{ii}} &= \ket{ii}, \\
\label{two_dicke2}    
    \ket{D_{ij}} &= \frac{1}{\sqrt{2}} (\ket{ij} + \ket{ji})~, \quad i \neq j~,
\end{align}
\noindent where $\{\ket{i} \in \mathbb{C}^{d}\}$.

Dicke states correspond to the block of Eq.(\ref{schur}) where $\bm{\lambda}$ is a partition of $N$ with exactly $d$ elements, and they span the so-called \textit{symmetric subspace} $\mathcal{S}_{N}^{(d)} \equiv \mathcal{S}((\mathbb{C}^{d})^{\otimes N})$, with dimension $\dim \mathcal{S}_{N}^{(d)}=\binom{N+d-1}{d-1}$. Notice that, in the case of qubits, i.e., $d=2$, the previous expression takes the simpler form: $\dim \mathcal{S}_{N}^{(2)} = N+1$. 

We conclude observing that the following series of inclusions holds true: symmetric states of $N$ qudits form a proper subset of permutationally invariant states of $N$ qudits, which in turn form a proper subset of general $N$ qudit states.
In the case of bipartite systems, these subspaces can be characterised more formally introducing the flip operator $F  = \sum_{ij} \ketbra{ij}{ji}$ and the projector onto the symmetric subspace $P_{s} = \frac{1}{2}(\mathds{1} + F)$. Hence, permutationally invariant states satisfy $F \rho F =  \rho$, while symmetric states are such that $ P_{s} \rho P_{s} = \rho$, as well as $F \rho F = \rho$. 

\section{Quantum correlations in symmetric states}
\label{sec:Correlations}

\noindent In this section, we present a comprehensive theoretical investigation of quantum correlations within the framework of symmetric states. We begin by examining entanglement, emphasizing the key differences between qubit and qudit systems. We then briefly introduce the concept of absolute separability in the symmetric subspace, followed by a discussion on the quantification of entanglement. Finally, we provide a concise overview of other forms of quantum correlations, including non-locality, steering, and non-classicality.

\subsection{Entanglement}
\label{sec:entanglement}

\noindent The reduced dimensionality of the symmetric subspace, compared to that of the full Hilbert space, implies that fewer degrees of freedom are required to describe such systems. For this reason, it is then natural to ask whether this simplification corresponds to an equivalent unravelling in the characterisation of entanglement in symmetric states. A complete answer to this question requires a careful analysis. In fact, while on the one hand symmetries often provide a nice framework to recast the original problem, on the other, many separability criteria happen to coincide for symmetric states \cite{toth2009entanglement}, thus requiring the use of new techniques.
Before examining the state of the art concerning the separability problem in the symmetric subspace, let us recall some basic properties of symmetric states. 

First, symmetric states are either fully separable or genuinely multipartite entangled (GME) \cite{ichikawa2008exchange}. As a result, symmetric separable pure states are of the form $\ket{\Psi}=\ket{e,e,\dots e }$, with $\ket{e}\in \mathbb C^d$. It follows that symmetric separable mixed states can be cast as
$\rho= \sum p_i \ket{e_i,e_i,\dots e_i}\bra{e_i,e_i,\dots e_i}$.
This simple yet powerful observation becomes especially significant when dealing with systems composed of a large number of parties. In such scenarios, one may be interested in estimating how many subsystems are entangled. This question leads to the introduction of two useful concepts: \textit{entanglement depth} and \textit{$k$-producibility}. Both capture intermediate levels of multipartite entanglement, accounting for situations in which a state is neither fully separable nor GME .
The \textit{entanglement depth} $n$ of an ensemble of $N$ parties indicates that the state involves genuine multipartite entanglement among at least $ n \leq N $ parties. By contrast, a state is said to be \textit{$k$-producible} if it can be decomposed into a product of states, each involving no more than $k$ entangled parties. Quite intuitively, these concepts are closely related: if a state has entanglement depth $n$, it is not $(n-1)$-producible.
In light of previous discussion, it is immediate to conclude that the entanglement depth of any entangled symmetric state of $N$ parties is given by $n=N$. However, it is important to note that this restriction does not apply to permutationally invariant states, where intermediate entanglement structures such as partial entanglement and $k$-producibility can indeed arise. Indeed, as soon as the states taken into consideration are not fully symmetric (i.e. their projection onto the symmetric subspace is not maximal) both these concepts of entanglement depth and k-producibility arise \cite{lucke_detecting_2014}.

\begin{figure}[h]
    \centering
    \includegraphics[width=0.5\linewidth]{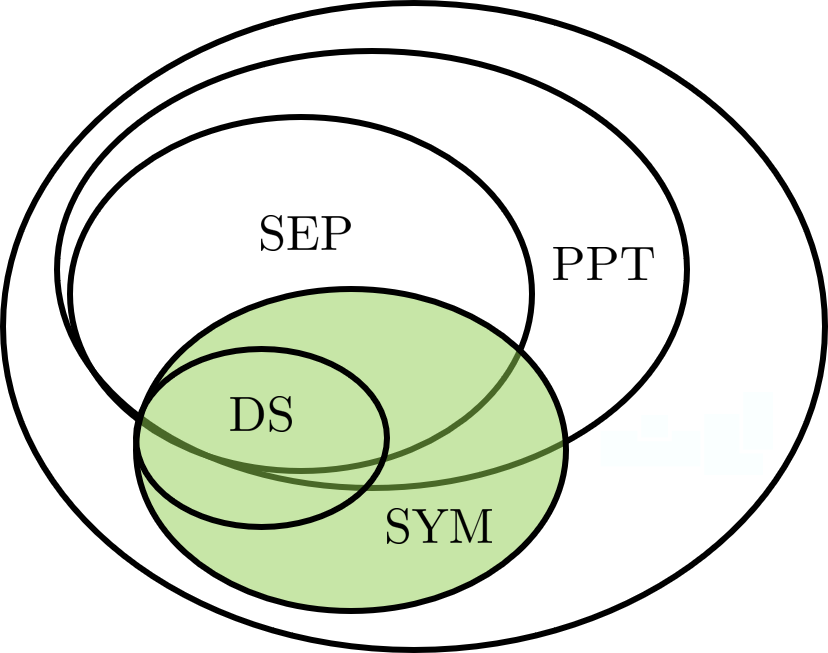}
    \caption{Pictorial representation of the convex set of quantum states. The symmetric subspace (SYM) and the subset of diagonal symmetric (DS) states are reported in green, along with the convex sets of separable states (SEP) and the set of states with positive partial transposition (PPT). }
    \label{fig:statespace}
\end{figure}

A second important observation about symmetric states arises from the fact that they can always be expressed in the Dicke basis. As a result, any $\rho \in \mathcal{S}_{N}^{(d)}$ can be decomposed as 
$\rho = \rho_{DS} + \rho_{CS}$, where $\rho_{DS}$ is a \textit{diagonal symmetric} state -- meaning that is diagonal in the Dicke basis --  while $\rho_{CS}$ contains the coherences between Dicke states. 
This decomposition is not only a practical starting point for analysing entanglement in symmetric states, but it is also conceptually significant.
Figure \ref{fig:statespace} provides a pictorial representation of the symmetric subspace in the generic bipartite case,  within the set of separable states and the set of PPT entangled states (i,e., entangled states that remain positive under partial transposition).
We start by analysing the structure of entanglement for qubit systems.

\subsubsection{Symmetric states of qubits}

\noindent The general form of an $N$-qubit symmetric state reads
\begin{equation}
\label{symm_qubit}
    \rho = \sum_{i,j=0}^{N} q_{ij}\ketbra{D^{N}_{i}}{D^{N}_{j}}~,
\end{equation}
\noindent where $q_{ij} \in \mathbb{C}$ and satisfy the normalisation constraint. A first restriction comes from considering DS states of the form
\begin{equation}
\label{DS_qubit}
\rho_{DS} = \sum_{i=0}^{N} q_{ii}\ketbra{D^{N}_{i}}{D^{N}_{i}}~,
\end{equation}
with $q_{ii} \geq 0, \sum_{i=0}^{N-1}q_{ii}=1$. For these states, it was first conjectured in \cite{wolfe2014certifying} and then independently proved in \cite{yu2016separability}  and \cite{quesada2017entanglement}, that positivity under partial transposition (PPT) is a sufficient condition for separability. Moreover, in the same works, it was proven that a DS state of $N$ qubits is PPT with respect to each partition if and only if it is PPT with respect to the largest partition, i.e., $\floor{N/2}:N-\floor{N/2}$, where $\floor{\cdot}$ denotes the floor function. Thus, there exists no DS PPT-entangled state (PPTES) of $N$-qubits and thus the separability problem in this subspace is completely solved. For higher dimensional systems, i.e. $N$-qudits, the situation changes and PPT is no longer sufficient for separability (see next subsection). However, as proved in \cite{PhysRevA.99.022309} using similar techniques, PPT is sufficient for separability for diagonal states w.r.t. the restricted basis  
\begin{equation}
    \ket{S_i^N} = \mathcal{N}_i  \sum_{ \mathbf{d}\cdot \mathbf{k} = i}{N\choose \mathbf{k}}^{1/2}\ket{D_\mathbf{k}^N}~,
 \label{Dicke_d_ basis}
\end{equation}
i.e., for states of the form $\rho_{DD} = \sum_{i=0}^{N(d-1)}q_{ii}\ketbra{S^N_i}{S^N_i}$ [cf. Eq.~\eqref{DS_qubit}], where $\mathbf{d} = (0,1,...,d-1)$ and $\mathcal{N}_i$ is the normalisation factor without simple closed formula. Intuitively, Eq.~\eqref{Dicke_d_ basis} represents a coherent superposition of qudit Dicke states with the same number of total excitations $i$ (for $N$ qubits $d=2$, $\ket{S_i^N} = \ket{D_i^N}$).

Extending the analysis to the whole symmetric subspace, one may ask whether the same result holds. While for $N=2$ the existence of PPTES is excluded by the PPT criterion, for $N=3$ it is proven that all PPT states are separable \cite{eckert2002quantum}. 
The first non trivial example of a PPTES of $N=4$ qubits is provided in \cite{tura_four-qubit_2012}, along with a half-analytical half-numerical method that allows to search for extreme states within the set of PPT states. Such method is based on the analysis of the triple $(r(\rho), r(\rho^{\Gamma_{1:3}}),r(\rho^{\Gamma_{2:2}}))$, where $r(\cdot)$ stands for the matrix rank and $\rho^{\Gamma_{m:n}}$ denotes the partial transposition with respect to the partition $m:n$. In particular, strong numerical evidence supports the claim that extremal PPTES for $N=4$ have ranks $(5,7,8)$, while other rank configurations are excluded. A similar analysis was extended in \cite{augusiak_entangled_2012} up to $N=23$ qubits, where recurring configurations of ranks for extreme PPTES were also identified.

\subsubsection{Symmetric states of qudits}
\noindent As one might expect, the analysis of entanglement in the symmetric subspace becomes significantly more challenging in high-dimensional systems compared to the qubit case. Nevertheless, when restricting to DS states, the characterization of entanglement can be approached by reformulating the original problem within a different framework. We begin by considering DS states of two qudits, whose explicit form is given by
\begin{equation}
    \label{DS_bi_qudit}
    \rho_{DS} = \sum_{i\leq j=0}^{d-1}p_{ij} \ketbra{D_{ij}}{D_{ij}}~,
\end{equation}
\noindent where $\{\ket{D_{ij}}\}$ are the two-qudit Dicke states of Eqs.(\ref{two_dicke1})-(\ref{two_dicke2}), and $p_{ij} \geq 0, ~\sum_{i\leq j}p_{ij} = 1$, due to normalisation. Particularly relevant is 
the direct sum structure arising in the partial transposition of bipartite DS states, \cite{tura2018separability}
\begin{equation}
\label{DS_PT_bi_qudit}
    \rho_{DS}^{\Gamma} = M_{d}(\rho_{DS}) \bigoplus_{i\neq j}\left(\frac{p_{ij}}{2}\right)~,
\end{equation}
where $M_{d}(\rho_{DS})$ is a $d\times d$ matrix, while $p_{ij}$ denotes a $1 \times 1$ block appearing with multiplicity $2$. As a consequence of Eq.~(\ref{DS_PT_bi_qudit}), there exists a one-to-one correspondence between  $\rho_{DS}$ (a matrix of dimension $d^2 \times d^2$) and  its associated matrix $M_{d}(\rho_{DS})$.  
This isomorphism allows to deduce the entanglement properties of the former state by examining properties of the latter matrix. Indeed, as shown in \cite{tura2018separability}, $\rho_{DS}$ is separable if and only if the matrix $M_{d}(\rho_{DS})$ is completely positive (CP), i.e., if there exists a $d \times k$ matrix $B$ such that $M_{d}(\rho_{DS}) = B B^{T}$, with $B_{ij} \geq 0 ~\forall i,j$. This result reveals a connection between two seemingly unrelated problems: determining separability within the diagonal symmetric subspace and verifying the complete positivity of a given matrix.

Completely positive matrices \cite{shaked2021copositive} find applications in a variety of contexts ranging from block designs \cite{hall1963copositive}
and Markovian models for DNA evolution \cite{kelly1994test}
to mathematical optimisation \cite{bomze2012think}
and machine learning \cite{kanamori2019model}.
Remarkably, such matrices form a convex cone, denoted as $\mathcal{CP}_{d}$, so that, in view of the previous result, assessing the separability of a bipartite DS state $\rho_{DS}$ is equivalent to checking the membership of its related matrix $M_{d}(\rho_{DS})$ to the cone $\mathcal{CP}_{d}$ of completely positive matrices. The investigation conducted in \cite{tura2018separability} highlighted also another interesting relation between bound entanglement in DS states and convex cones. Indeed, the cone of completely positive matrices satisfies the inclusion $\mathcal{CP}_{d} \subseteq \mathcal{DNN}_{d}$, where $\mathcal{DNN}_{d}$ is the convex cone of doubly nonnegative (DNN) matrices, that is those matrices that are both positive semidefinite and entry-wise nonnegative. Curiously, while for $d \leq 4$ the two cones coincide, for $d \geq 5$ the inclusion is strict, i.e., $\mathcal{CP}_{d} \subset \mathcal{DNN}_{d}$, and there exist matrices such that $M_{d}(\rho_{DS}) \in \mathcal{DNN}_{d} \setminus \mathcal{CP}_{d}$. In  \cite{tura2018separability} it was proven that such matrices correspond to PPT-entangled DS states, and vice versa, providing also two explicit examples for $d=5$ and $d=6$. Due to the peculiar relation between $\mathcal{CP}_{d}$ and $\mathcal{DNN}_{d}$, DNN but not CP matrices can only occur when $d\geq 5$. As a consequence, PPT-entangled DS states of two qudits cannot exist for $d \leq 4$ and, in this case, the PPT criterion becomes necessary and sufficient for separability. 

\begin{figure}[h]
    \centering
\includegraphics[width=0.9\linewidth]{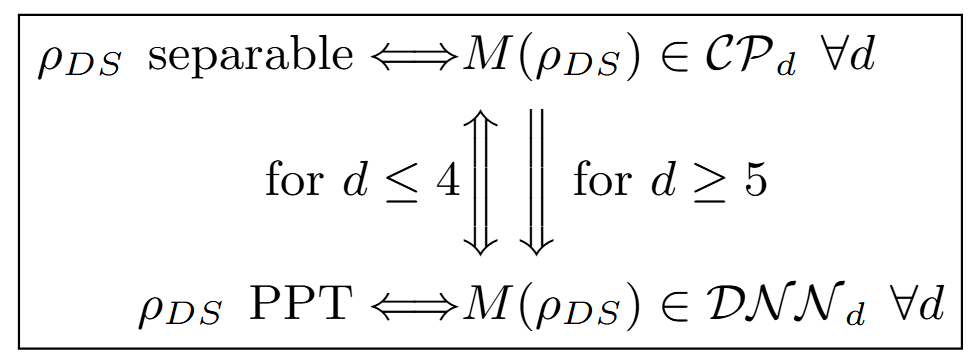}
\caption{Relationship between the properties of a two-qudit DS state $\rho_{DS}$ and its associated matrix $M_d(\rho_{DS})$. For every dimension $d$, deciding the separability of $\rho_{DS}$ is equivalent to checking the complete positivity of $M_d(\rho_{DS})$. For $d\leq 4$, PPT is both a sufficient and necessary criterion for separability, while for $d \geq 5$ it is a necessary (but not sufficient) condition due to the existence of DNN but not CP matrices.}
    \label{fig:DS_Summary}
\end{figure}

We stress that, while the DNN condition can be easily checked for a generic matrix $M_{d}(\rho_{DS})$, the same is no longer true for the CP condition and indeed this task is known to be an NP-hard problem \cite{dickinson2014computational}. This observation should not come as a surprise, being a consequence of the computational complexity of the original separability problem. Nevertheless, making use of a correspondence between matrices and graphs, there exist several criteria that allow to check for complete positivity (see \cite{shaked2021copositive} and references therein). Such criteria are particularly suitable for numerical programming, especially in light of the fact that the dimension of a matrix $M_{d}(\rho_{DS})$ is greatly reduced as compared to the one of its associated quantum state $\rho_{DS}$, thus making this correspondence a valuable tool for the separability problem in the DS subspace. A brief summary of these results can be found in Figure \ref{fig:DS_Summary}.

The above analysis can be extended beyond the bipartite case showing that it is still possible to establish a correspondence between a multipartite DS state, $\rho_{DS}^N \in \mathcal{S}_{N}^{(d)}$, and a matrix $\tilde{M}_{d} (\rho_{DS})$ arising from the partial transposition w.r.t. the largest partition, although, in principle, all possible partitions must now be considered. The structure of such matrix is, in general, quite involved. Nevertheless, there is strong analytical and numerical evidence that positivity w.r.t. the largest partition implies also positivity w.r.t. to any other partition, analogously to the case of $N$-qubit DS states \cite{PrivateCommunicationJordi}. However, in contrast to the qubit case, already for three qutrit DS states there exist PPTES \cite{PrivateCommunicationNechita}.

\begin{figure}[h!]
    \centering
    \includegraphics[width=1\linewidth]{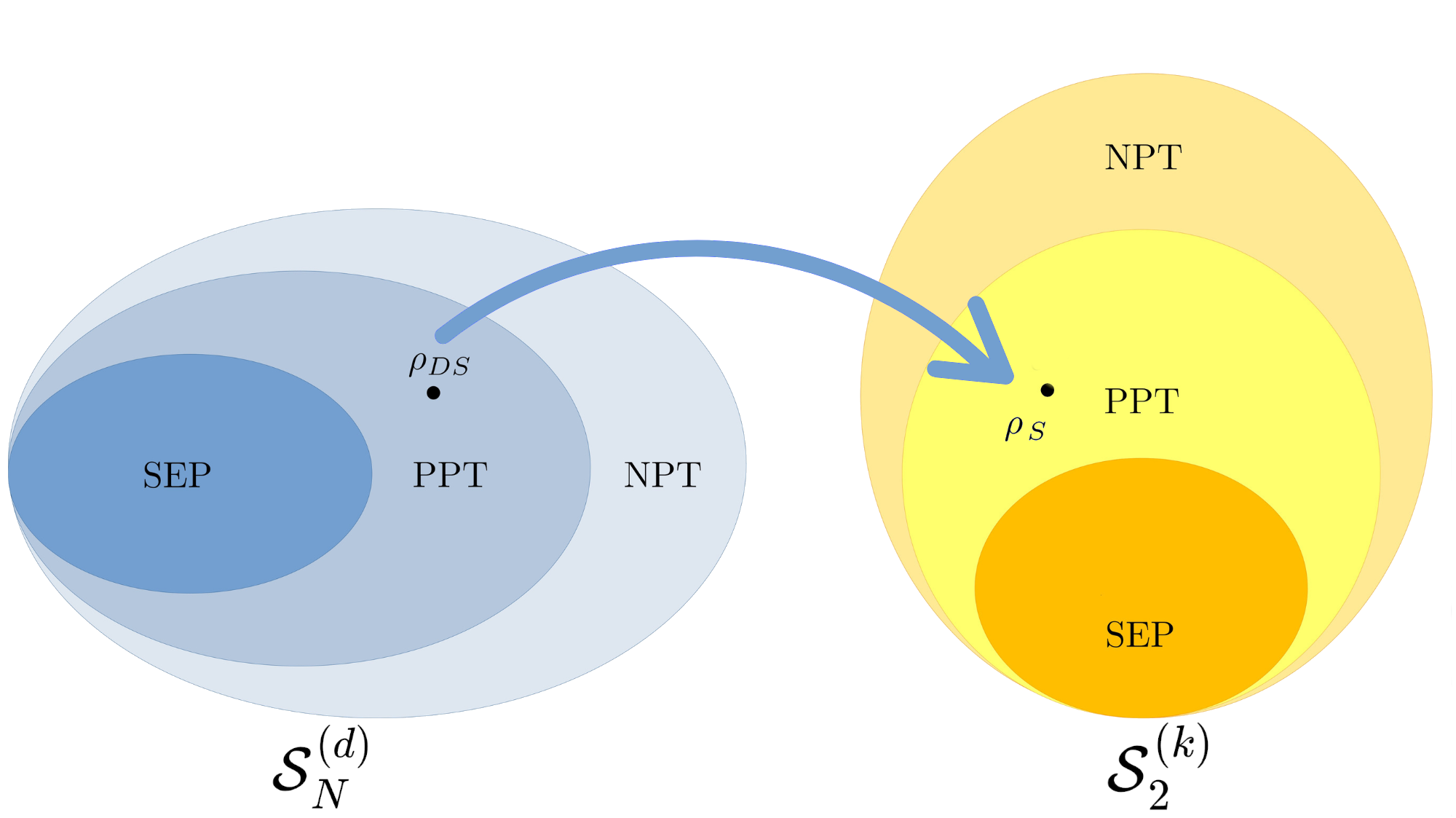}
    \caption{Mapping a multipartite DS state $\rho_{DS} \in \mathcal{S}_{N}^{(d)}$ to a bipartite symmetric state $\rho_{S} \in \mathcal{S}_{2}^{(k)}$ with $k = \binom{N/2+d-1}{d-1}$. Here $N$ is assumed to be an even number.}
    \label{fig:embeddingA}
\end{figure}

We conclude this section with a brief survey on entanglement in symmetric states of qudits, beyond the subset of DS states. Starting from the bipartite case, the first examples of PPT-entangled symmetric states were reported in \cite{toth2009entanglement}, where two methods to generate such states were provided: one relies on embedding entangled states within a Hilbert space of larger dimension, ensuring that both symmetry constraints and PPT condition are satisfied; alternatively, a half-analytical half-numerical method was proposed where bipartite PPT-entangled symmetric states are constructed starting from an $N$-qubit symmetric state which is PPT w.r.t. the partition $\frac{N}{2}:\frac{N}{2}$, while NPT w.r.t. the one of the others (here, $N$ is an even number). With the aid of this technique, the existence of a two-qutrit symmetric bound entangled state was reported. Remarkably, a similar approach, where a multiparty symmetric $N$-qudit DS state is embedded into a bipartite symmetric state of higher local dimension can be adopted \cite{PrivateCommunicationJordi} to relate the entanglement properties of multiparty systems to those of bipartite ones (see Figure \ref{fig:embeddingA}). This approach also has the advantage of reducing the computational cost needed to assess the separability properties of DS states.

The complete characterisation of PPTES in the symmetric subspace of two qutrits was provided in \cite{romero2025preparation}, focusing on the class of \textit{edge} PPT-entangled states. We recall that a (bipartite) PPTES $\delta$ is said edge if, for every product vector $\ket{e,f}$ and any $\varepsilon >0$, $\delta - \varepsilon \ketbra{e,f}{e,f}$ is not a PPTES \cite{lewenstein2000optimization}. From their definition, it is evident that edge states lie at the border between the sets of PPT and NPT states. In \cite{romero2025preparation}, it was proven that any two-qutrit symmetric edge state takes the form
\begin{equation}
\rho=\rho_{DS}+\sum_{\substack{0\leq i \leq 2\\ i\neq j \leq k }}\lambda_i\ketbra{D_{ii}}{D_{jk}}~,
\end{equation}
with $\sum_{i} (\lambda_i/\lambda'_i)>1$. Here  $\lambda'_i$ is defined as the maximum value such that $\rho_{DS}+\lambda'_i\ketbra{D_{ii}}{D_{jk}}$ is a valid PPT state (i.e., both the state and its partial transposition are positive semidefinite). Notice that when the ranks of the state and its partial transposition are such that $r(\rho,\rho^{\Gamma})=(5,7)$, the corresponding state is edge. Since edge states are extremal within the set of PPT states, understanding their structure is sufficient to characterize the entire PPT set. We conclude this section by noting that, making use of the PPT symmetric extension \cite{doherty2004complete}, it is possible to generate multipartite PPT-entangled symmetric states for arbitrary number of parties \cite{romero2025preparation}.

\subsubsection{Absolute separability}
\noindent  We present here recent results concerning absolute separability (AS) within the symmetric subspace. AS denotes the set of separable quantum states that remain separable under the action of any global unitary transformation. Consequently, states of the  AS set depend solely on their spectrum (i.e., eigenvalues) and, for this reason, this property is also referred to as ``separability from spectrum". Progress in characterising the extremal points of the convex AS set has been quite limited. However, a renewed interest has recently emerged \cite{Halder_2021,champagne2022spectral,Serrano2024,louvet2025nonequivalence,abellanet2024improving}, partly driven by the potential role of AS as a foundation for a restricted resource theory of entanglement. Since unitary transformations are the primary way of generating entanglement -- whether through quantum circuits or quenches --  AS states cannot be transformed into entangled one by unitary evolution. Sufficient conditions for absolute separability using the inverse of linear maps have recently been proposed \cite{abellanet2024improving} which involved only one or two eigenvalues, however, the complete characterization of the AS set remains an open problem.  A related issue is the so-called absolute PPT (APPT) problem, which concerns the convex set of states that remain PPT under any unitary transformation. In contrast to the AS set, the APPT set is completely characterized in arbitrary dimensions using a hierarchy of linear matrix inequalities \cite{Hildebrand2007positive}, though closed conditions cannot be obtained from them. Whether the AS and APPT sets coincide remains an open conjecture in entanglement theory.

These concepts can be naturally adapted to the symmetric subspace by defining the set of symmetric absolutely separable (SAS) and the set of symmetric absolute PPT (SAPPT). Formally, a symmetric state $\rho_S$ belongs to the SAS set if and only if  $\rho'_S=U_S\rho_S U_S^{\dagger}$ remains fully separable (respectively, fully PPT) for all symmetric unitaries $U_S$.

Characterizing the SAS set—much like in the general AS case—is a highly non-trivial task. Tight bounds on this set are currently known only for specific bipartite systems \cite{champagne2022spectral,Serrano2024}. In \cite{louvet2025nonequivalence} explicit examples demonstrating that the SAS and SAPPT sets are not equivalent are provided. For multipartite systems, 
$N=3,d=2$, SAS bounds are provided in \cite{abellanet2024improving}.

\subsubsection{Quantification of multipartite entanglement}

The quantification of entanglement in multipartite quantum systems remains a central challenge in quantum information theory, particularly in the case of mixed or many-body states. To this end, the main tool is represented by an \textit{entanglement measure}, i.e., 
a functional $\mathcal{E}$ satisfying the following properties (see \cite{Plenio2014}):
\begin{itemize}
    \item[1.] $\mathcal{E}: \mathcal{B}(\mathcal{H}) \rightarrow\mathbb{R}^{+}_{0}$ associates any state $\rho \in \mathcal{B}(\mathcal{H})  $ to a nonnegative number $\mathcal{E}(\rho) \geq0$.
    \item[2.] $\mathcal{E}(\rho)=0$ if and only if $\rho$ is a separable state
    \item[3.] $\mathcal{E}(\rho)$ does not increase on average under LOCC, i.e., 
    $\mathcal{E}(\rho) \geq \sum_{i} \mathcal{E}(K_{i} \rho K_{i}^{\dagger})$~,
    where the $K_{i}$'s are the Kraus operators associated to the LOCC protocol.
    \item[4.] For pure states, i.e., $\rho=\ketbra{\psi}{\psi}$, $\mathcal{E}$ is equal to the entropy of entanglement.
\end{itemize}
While for bipartite pure states all entanglement measures coincide, allowing to define with no ambiguity maximally entangled states, moving to multipartite states this is no longer the case and states that maximize one entanglement measure do not necessarily maximize another. 
Among the various entanglement measures, the geometric measure (GM) of entanglement \cite{shimony1995degree,wei2003geometric} stands out for its simple definition and operational significance, being related to state discrimination under LOCC \cite{hayashi2006bounds}. For pure states, the GM quantifies entanglement by measuring the minimal distance between a given state and the set of fully separable states (Sep). 
\begin{equation}
    E_G(\ket{\psi})=1-\max_{\ket{\phi}\in \text{Sep}} |\langle\phi|\psi\rangle|^2.
\end{equation}

The above expression requires an optimization procedure over the separable set and, for this reason, even for pure states, calculating the geometric measure is generally a hard task~\cite{de2008tensor}.
Quite intuitively, extending the geometric measure to mixed states is even more challenging, since it requires a convex roof construction.
\begin{equation}
    E_G(\rho)=\min_{p_k,\ket{\psi_k}} \sum p_k E_G(\ket{\psi_k}),
\end{equation}
where the minimization is taken over all convex decompositions of $\rho$.
This is computationally intractable in general, making exact evaluations unfeasible except in special cases. One such case arises in the study of symmetric states, where remarkable simplifications occur.

A key breakthrough was the realization that for multipartite symmetric states, the optimization in the GM can be restricted to symmetric separable states. Initially, it was shown that one can always choose a closest separable state that is symmetric \cite{hayashi2008entanglement}. Later, this result was strengthened, proving that the closest separable state is necessarily symmetric \cite{hubener2009geometric}, thus confining the variational problem to a much smaller subspace. 

The GM has been employed as a tool for many things, specially to study multipartite entanglement  and to identify candidates for maximally entangled symmetric states and to study their robustness \cite{g_hne_2008}. In particular, \cite{martin2010multiqubit} analytically derived an upper bound for the GM of pure $N$-qubit symmetric states. Additionally, states exhibiting high geometric entanglement were identified based on an intriguing connection with the Majorana representation, a geometric tool that maps symmetric states to configurations of points on the Bloch sphere. This correspondence was investigated also in \cite{aulbach2010maximally}, where pure symmetric $N$-qubit states with high GM were numerically identified for systems up to $N=20$. The relationship between the Majorana representation and the GM for symmetric states was further explored in \cite{markham2011entanglement}, where it was used to demonstrate the equivalence between various entanglement measures. More recently, an alternative approach, based on a linear mapping between symmetric subspaces, has been proposed to estimate the amount of entanglement in pure symmetric $N$-qubit states \cite{marconi2025entanglement} and provide some lower bounds on the GM of a multipartite state. For 3 qubits highly symmetric states, such as the GHZ-symmetric and W states (and mixtures of those), the 3-tangle measure\cite{Coffman_2000} has also been computed \cite{Lohmayer_2006, Eltschka_2012}.

A different approach in the quest for maximally entangled states is to consider quantum states whose one-body marginal are maximally mixed. These states have been studied in several papers (see, e.g., \cite{goyeneche2014genuinely} and references therein). A first step towards this characterisation was provided in \cite{arnaud2013exploring}, where it was proven that $N$-qubit symmetric states cannot have maximally mixed $k$-qubit reductions with $k>1$. Remarkably, this problem is also related with the fact that absolutely maximally entangled (AME) states cannot exist in the $N$-qubit symmetric subspace. Further, in \cite{baguette2014multiqubit}, it was found a general criterion to decide whether $N$-qubit symmetric states are maximally entangled in the above sense. Interestingly, this problem appears to be related to the entanglement classification via stochastic LOCC (SLOCC), a problem which was completely solved in \cite{bastin2009operational}.

\subsection{Nonlocality}
Nonlocality is a fundamental phenomenon in quantum mechanics, where local measurements on a composite system distributed among distant parties can exhibit correlations which are impossible to reproduce by any local deterministic strategy assisted by shared randomness \cite{bell_einstein_1964,brunner_bell_2014}. This feature makes nonlocality a valuable resource for various information-theoretic tasks, ranging from secure key distribution \cite{ekert1991quantum,acin2007device} to certify quantum randomness generation \cite{pironio2010random}.
In order to detect nonlocality, one typically employs Bell inequalities, consisting of linear inequalities expressed in terms of correlators, i.e., expectation values of the local measurements performed by observers. 
Similarly to the case of entanglement witnesses, also Bell inequalities are endowed with a geometrical interpretation. In fact, it is possible to show that the set of local correlations, $\mathcal{L}$, resulting from the requirement that all the parties can communicate using only local strategies, is a polytope, i.e., a compact, convex set, with a finite number of extreme points. Hence, as a consequence of the Hahn-Banach theorem, Bell inequalities can be interpreted as hyperplanes separating any point outside $\mathcal{L}$ from its interior. Formally, introducing the Bell operator $\mathcal{B}$ corresponding to a given Bell inequality, condition $\mbox{Tr}(\mathcal{B} \rho)<0$ signals the presence of nonlocal correlations in the state $\rho$. Moreover, analogously to the case of optimal entanglement witnesses, tight Bell inequalities can be regarded as those hyperplanes which are tangent to the local polytope $\mathcal{L}$. The characterisation of such Bell inequalities is particularly relevant, since they provide a minimal representation of $\mathcal{L}$, meaning that any other Bell inequality can be written as a convex combination of them. However, the intricacy of this task makes this extremely difficult to solve already in the case of a small number of parties and measurement settings \cite{sliwa2003symmetries} and indeed such task has been proved to be an NP-hard problem \cite{brunner_bell_2014}. One intuitive argument to understand the origin of this complexity is that, when dealing with $N$ parties, the construction of a Bell inequality typically requires to address the full-order correlators, making this problem easily unfeasible.

\noindent In light of this discussion, one might consider a simplification of the above problem where only low-order correlators are considered. Indeed, this idea was originally proposed in \cite{tura2014detecting}, where the authors focused on an approach where only one- and two-body correlators are considered, requiring them to remain invariant under any permutation of the parties. As a consequence, $\mathcal{L}$ is projected onto a simpler object, dubbed the permutationally invariant polytope, whose facets correspond to Bell inequalities of the form \cite{tura2015nonlocality}
\begin{equation}
\label{bellfamily}
 \beta_{\emph{cl}} + \alpha \mathcal{C}_{0}  + \beta \mathcal{C}_{1} + \frac{\gamma}{2}   \mathcal{C}_{00}  + \delta \mathcal{C}_{01}  + \frac{\epsilon}{2}   \mathcal{C}_{11} \geq 0~,
\end{equation}
where $\alpha,\beta,\gamma,\delta,\epsilon \in \mathbb{R}$, $\beta_{\emph{cl}} \in \mathbb{R}$ is the so-called classical bound of the associated Bell inequality and
\begin{equation}
\mathcal{C}_{r} = \sum_{i=0}^{N-1}  \langle \mathcal{M}^{(i)}_{r} \rangle~, \quad
 \mathcal{C}_{rs}  = \sum_{{i \neq j=0}}^{N-1}   \langle \mathcal{M}^{(i)}_{r}\mathcal{M}^{(j)}_{s} \rangle~,
\end{equation}
 are the one- and two-body permutationally-invariant correlators, respectively. Here, $\langle \mathcal{M}^{(i)}_{r} \rangle $ denotes the expectation value of the measurement operator for the $i$-th party, corresponding to the measurement setting $r \in \{0,1\}$.
 Notice that, although the Bell inequalities in Eq.~(\ref{bellfamily}) are invariant under any permutation of the parties, the corresponding Bell operator, as well as the nonlocal states violating such inequalities, do not necessarily share the same symmetry. Nevertheless, in \cite{tura2014detecting}, it has been shown that for a suitable choice of the parameters, the Bell inequality of Eq.~(\ref{bellfamily}) is violated by $N$-qubit Dicke states for any value of $N$. Moreover, in \cite{tura2015nonlocality}, the authors investigated the maximal violation of such Bell inequalities, providing strong numerical evidence that, when all the parties perform the same pair of measurements, this occurs for a superposition of $N$-qubit Dicke states of the form $\ket{\psi} = \sum_{k=0}^{N} c_{k} \ket{D^{N}_{k}}$ with $c_{k} \in \mathbb{R}$. The non-local behaviour of symmetric systems of $N$ qubits was explored also using alternative techniques, e.g., in \cite{wang2012nonlocality,wang2013nonlocality}, where the authors exploited the Majorana representation of a symmetric state as well as semidefinite programming techniques, to construct Bell inequalities for multipartite W and Greenberger-Horne-Zeilinger (GHZ) states.
 Focusing on DS states of qubit systems, in \cite{quesada2017entanglement} it was shown, using only one- and two-body Bell inequalities, that there exist states that are PPT w.r.t. to one partition but are nevertheless non-local, thus providing a counterexample to the celebrated Peres conjecture \cite{peres1996separability}. 

In \cite{FadelQuantum2018}, the authors explored a  connection between nonlocality and temperature in a system of spins with infinite-range interactions. Interestingly, they report the violation of a multipartite Bell inequality from thermal equilibrium up to a finite critical temperature. 

Along a similar line of research, \cite{marconi_robustness_2022} explores the behaviour of nonlocality in a many-body open quantum system composed of $N$ qubits, interacting with an external environment. Assuming a model whose stationary solutions correspond to permutationally invariant states, the authors provide evidence that nonlocal correlations are robust and can survive for a certain time, also when noise is taken into account. 
A similar problem is considered in \cite{guo_detecting_2023}, where the authors propose a general framework to obtain experimentally-feasible high-order PI correlators for non-Gaussian states, which are robust to noise and display a better sensitivity.

Moving to the case of qudit systems, in \cite{muller-rigat_inferring_2021} a new approach to derive Bell inequalities for high-dimensional states was proposed. In particular, the authors reported the construction of an algorithm whose complexity grows independently on the number of parties and allows to retrieve the family of permutationally invariant Bell inequalities found in \cite{tura2014detecting} while enabling an extension to the case of more measurement settings and outcomes. Similar three-outcome Bell inequalities tailored to many-body spin-1 systems were studied in a series of works~\cite{muller2024three,aloy2024deriving, aloy_bell_2024} which also explored their implications in quantum chaos, squeezed states, and dimension witnessing. On the other hand, in \cite{wagner_bell_2017} such inequalities were extended to arbitrary number of measurement settings.   

\subsection{Other quantum correlations}
Beyond entanglement and nonlocality, quantum systems can exhibit a rich variety of other quantum correlations. One notable example is quantum steering, a concept introduced by Schrödinger in his seminal 1935's paper \cite{schrodinger1935discussion}, where the term referred to the ability of one party of an entangled system to influence the state of the other by performing local measurements on her own subsystem. Remarkably, this effect cannot be accounted for by classical theories or local hidden variable models. In spite of its relevance, a formal definition of quantum steering was given only almost 80 years later in \cite{wiseman2007steering}. There, the authors proved that the set of steerable states is a strict subset of the set of entangled states and a strict superset of the set of nonlocal states. As such, quantum steering can be seen as an intermediate form of correlation between entanglement and Bell nonlocality.
Focusing on permutationally invariant systems, \cite{pan2021nonlinear} established a steerability criterion for two-qubit systems, while \cite{wasak2018bell} proposed an approach to address the relation between steering and Bell correlations. Furthermore, steering conditions for symmetric $N$-qubit states were initially explored for the case of $N = 3$ \cite{anjali2022geometric}, and later generalized to arbitrary $N$ \cite{divyamani2023canonical} making use of the so-called steering ellipsoids. This line of work also revealed a connection between steering and the classification of symmetric states via SLOCC.
Moreover, steering has been observed in two spatially-separated Bose-Einstein condensates~\cite{he2013einstein, FadelScience2018,colciaghi2023einstein} from witnesses based on uncertainty relations which are specially tailored to such many-body systems~\cite{reid1989demonstration, vitagliano2023number}. Finally, in~\cite{Yadin2021} steering is proven to be a resource for metrological applications, in which symmetric states play a central role as discussed later in this review.

Another benchmark of quantum behavior is \textit{nonclassicality}, a concept originally introduced in quantum optics to describe states of light that cannot be explained within the framework of classical electromagnetic theory. In this setting, coherent states — which minimize the uncertainty principle — are considered the most ``classical" among quantum states. Consequently, a quantum state is said nonclassical if it cannot be expressed as a statistical mixture of coherent states. More formally, this condition is equivalent to the negativity of the celebrated P-function, proposed by Sudarshan and Glauber as a key signature of nonclassical light \cite{sudarshan1963equivalence,glauber1963coherent}. An analogous characterisation can be given in terms of the Wigner function \cite{wigner1932quantum}, whose negativity can be regarded as an alternative indication of nonclassicality \cite{kenfack2004negativity}. When dealing with spin systems, coherent states of light are replaced by spin-coherent states and a corresponding Wigner function can be defined \cite{klimov2017generalized}. Leveraging on the correspondence between spin states $\ket{J,M}$ and $N$-qubit Dicke states $\ket{D_{k}^{N}}$ with $N=2J$ and $k=N/2-M$, the negativity of the Wigner function for spin-$J$ systems has been examined in \cite{davis2021wigner} for several pure symmetric states. More recently, in \cite{denis2024polytopes}, the negativity of Wigner function for spin systems was connected with the problem of absolute separability for symmetric states. The theory of non-negative polynomials has been also used to reveal nonclassicality in both photon and spin systems~\cite{PhysRevLett.134.030201}.

We conclude this section by highlighting an alternative approach to characterize the nonclassicality of quantum states, which relies on constructing indicators of \textit{quantumness} \cite{goldberg2020extremal}. An early contribution in this direction was made in \cite{giraud2010quantifying}, where a distance-based measure for spin systems was introduced to identify the most nonclassical states, referred to as Queens of Quantum. A related class of states, termed Kings of Quantum, was later identified in \cite{bjork2015extremal} using a different method based on the multipole expansion of the density matrix. More recently, \cite{goldberg2022quantumness} proposed an entanglement-based measure for symmetric states, offering yet another perspective on quantifying the most quantum states within this set. Notably, in all these studies, the Majorana representation of symmetric states not only reduces the complexity of the problem but also provides valuable geometric insight that guides the search for extremal quantum states.

\section{Certification of symmetric states}
\label{sec:Characterization}

\noindent Quantum phenomena do not occur in a Hilbert space, they occur in the laboratory~\cite{Peres2002}. Thus, it is relevant to devise protocols to \textit{witness} permutation invariant (PI) and symmetric states from experimental data. We start by addressing quantum state tomography and detailing how symmetry allows to reduce partially its computational cost. Next, we review non-tomographic protocols to verify the preparation of symmetric states, including device-independent approaches such as self-testing. Finally, we deal with entanglement certification, highlighting the close relation between the structure of the states and of their corresponding witnesses.

\subsection{Quantum state tomography}

\noindent The density matrix of a system can be inferred from experimentally-available probes via a process known as quantum state tomography (QST)~\cite{paris2004quantum}. In the multipartite scenario, QST protocols become quickly unfeasible, partly due to their underlying combinatorial complexity. For instance, the 8-qubit state generated in the ion experiment shown in \cite{Haffner2005} took several weeks to be fully reconstructed~\cite{Gross2010}. 

Permutation invariant (PI) QST aims at inferring soley the PI sector of the state, which constitutes a good approximation to the whole state in many relevant cases~\cite{ Toth2010, Moroder2012}. In addition, the PI part of a density matrix contains essential information about its entanglement content~\cite{Gao2014}. Let us consider a generic $N$-qubit state $\rho$. Its projection to the PI subspace, $\rho_{\rm PI} $ can be obtained by applying the averaging map:
\begin{equation}
    \rho_{\rm PI} = \frac{1}{N!}\sum_{\pi\in\mathcal{G}_N} P_\pi \rho P_\pi^\dagger~,
\end{equation}
where $\{P_\pi\}$ is the set of permutation matrices and $\mathcal{G}_N$ is the symmetric group as defined in Section~\ref{sec:Symmetry}. We remind the reader that, by virtue of the Schur-Weyl duality, the PI state can be block-diagonalized. In the case of qubits, such as e.g. spin-$1/2$ particles, (cf. Eq.~\eqref{eq:PIdecom} in Section~\ref{sec:Symmetry}) 
\begin{equation}
    \rho_\mathrm{PI} \cong \bigoplus_{J=J_{\min}}^{N/2} \rho_J,
\label{eq:SWJ}
\end{equation}
where the $J=N/2$ sector covers the fully symmetric subspace. More generally, the $J$-th sector can be interpreted as a magnified spin of length $\sqrt{J(J+1)}$, with $J(J+1)$ being the quantum number associated to the total spin $\mathbf{J}^2 = J_x^2 + J_y^2 + J_z^2$. 
In the corresponding irreducible representation, this subspace is spanned by the vectors $\{\ket{J,M}\}_{M \in \{ -J, -J+1,..,J\}}:=\mathcal{C}$, where $M$ is the quantum number associated to the spin projection $J_z$. In the simple case of two qubits ($N= 2$), the space decomposes into symmetric and antisymmetric parts, corresponding to the triplet $J=1$ and singlet $J=0$, respectively. Beyond bipartite systems ($N>2$), one may further coarse-grain the space by identifying states that can be related by permutations of the parties, i.e., with the same total spin $\mathbf{J}^2$ and projection $J_z$, to construct the Dicke basis $\mathcal{C}$. Note how the dimension of this coarse-grained space grows only linearly with the system's size $N$, which contrasts with the exponential scaling for generic $N$-partite states. Hence, we also expect a favorable scaling in the number of measurement settings required to fully specify the PI state. Indeed, in \cite{Toth2010} it was proven, using counting arguments, that only ${N-2 \choose N} \sim \mathcal{O}(N^2) $ binary local measurement settings are sufficient to fully describe $\rho_{\rm PI}$. These correspond to averaged $K$-body correlators of the form
\begin{equation}
   \langle(A^{\otimes K}\otimes \mathds{1}^{N-K}_2)_{\mathrm{PI}}\rangle~,
   \label{eq:PImeas}
\end{equation}
where $(\cdot)_{\mathrm{PI}} = \frac{1}{N!}\sum_{\pi\in \mathcal{G}_N} P_\pi\cdot P_\pi$, $A$ is an individual observable such as $A^2 = \mathds{1}_2$ (e.g. Pauli matrices), $\mathds{1}_2$ is the qubit identity matrix and $K \in \{ 0,1,..,N-1\}$. The observable $A$ may be optimised to minimise the error to statistically infer the state from its expectation values, based on a finite number of copies available to the experimentalist.  

To conclude, PI QST enables reconstruction of PI states (and, in particular, of symmetric states) for larger number of qubits $N$. The advantages of this approach together with compressed sensing ~\cite{Gross2010} are acknowledged in \cite{Schwemmer2014} by analyzing a photonic 6-qubit Dicke state. Besides PI QST, additional techniques tailored to symmetric states include the use of optimal mutually unbiased bases~\cite{Klimov2013}. Other recent advances in the field comprise classical shadows~\cite{Aaronson2017} and machine learning-assisted QST~\cite{Torlai2018} as well as a novel approach, based on the symmetrization of observables, which allows for a significant reduction of the sample complexity \cite{Zhang_2024}.

Nevertheless, QST is inherently a resource-intensive task and it is severely limited by the system's size, even when symmetries are exploited. For instance, the number of samples sufficient to infer $K$-body correlations does not scale well with $K$ [more exactly as $\Theta(3^K)$]. Consequentially, it prompts the need of non-tomographic methods to certify the preparation of states under symmetry, possibly using only partial information. We address these approaches below. 

\subsection{Verification}

\noindent Full tomography may not be necessary to extract useful features of a quantum state, such as its symmetry or entanglement content. For example, consider the projector onto the totally symmetric sector, $P_S = \sum_{k=0}^{N}\ketbra{D_k^N}{D_k^N}$. If its expectation value is sufficiently close to $1$, we can conclude the underlying state is almost fully symmetric. Conversely, a single observable can be enough to certify entanglement -- specifically, an entanglement witness $W$. This highlights an important insight: certain experimental outcomes provide more information than others. One can take advantage of this aspect to develop novel approaches to efficiently verify the preparation of a target state within a given confidence level, for example, by designing optimal measurement routines to estimate the fidelity between the prepared state and the pure target state. However, unlike QST, if such verification test fails, it yields no further information about the nature of the prepared state.

To date, optimal verification protocols exist for states that admit a stabilizer formalism ~\cite{Dangniam2020, Kalev2019}, such as graph states~\cite{Li2023} or GHZ states~\cite{Li2020}. As an illustration, for 2 qubits, the GHZ state, $\ket{\mathrm{GHZ}} = (\ket{00} + \ket{11})/\sqrt{2}$, is the unique simultaneous $+1$ eigenvector of $\mathcal{T} = \{X\otimes X, Z\otimes Z \}$, where $\mathcal{T}$ is the so-called stabilizer of GHZ. Note how elements in $\mathcal{T}$ pairwise commute and are product operators. Thus, to assess how far is our actual state from the GHZ, we can measure $\mathcal{T}$ in order to discriminate the target state from imperfect preparations. The number of Pauli measurements needed to estimate the fidelity between the target and the actual state can be reduced by a factor corresponding to the global Hilbert's space dimension compared to QST techniques~\cite{Flammina2011}. \\

\begin{figure}
    \centering
    \includegraphics[width = 0.60\columnwidth]{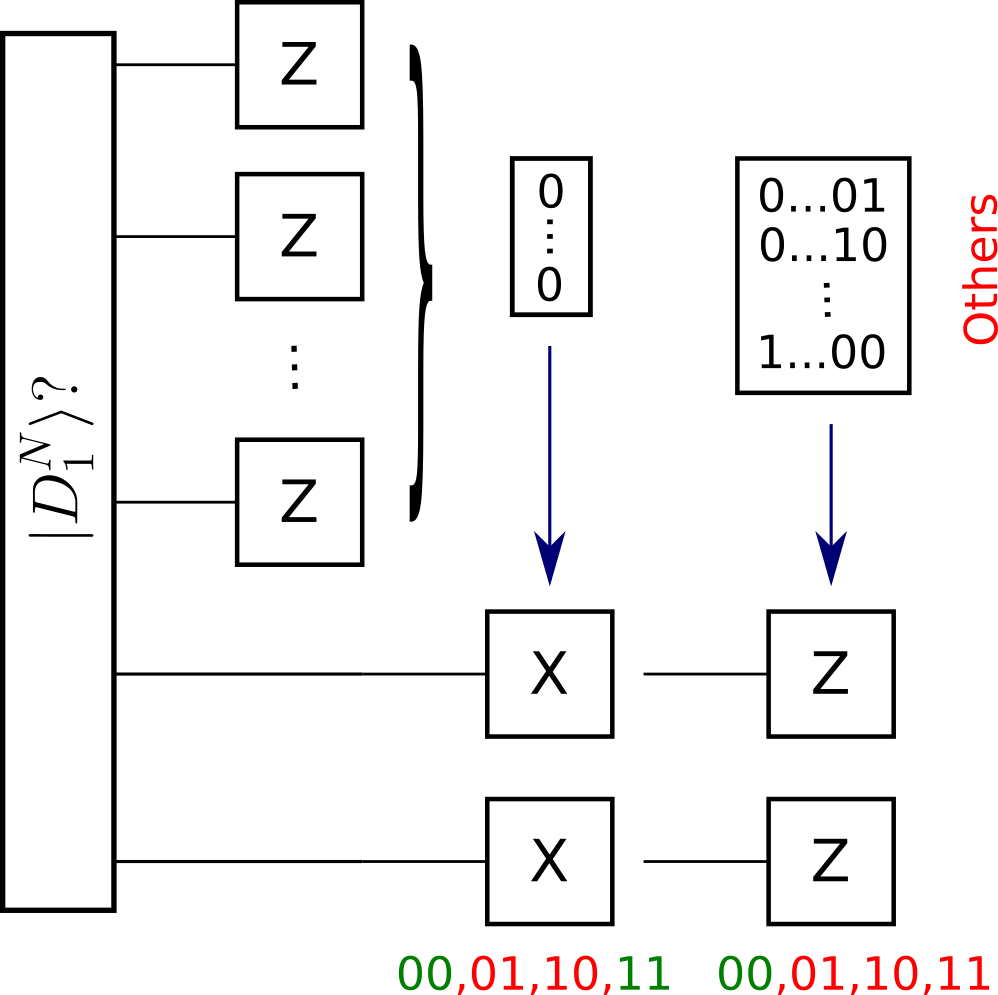}
    \caption{Verification scheme for the $\ket{D^N_1}$ state via the adaptive strategy reported in \cite{Pallinster2018}. The Pauli $Z$ operator is measured in $N-2$ parties. If it yields more than 1 excitation, we stop the procedure as one can readily deduce that the initial state was not $\ket{D^N_1}$. When only 1 excitation is observed, we measure $Z$ in the remaining 2 particles, while if we get $0...0$, we measure $X$. The outcomes of such measurements, will determine whether the state was $\ket{D^N_1}$ (red) or if it passes the test (green). Source: inspired from \cite{Pallinster2018}.}
    \label{fig:DickeVeri} 
\end{figure}

\noindent \textbf{Verification of Dicke states}.-- In \cite{Liu2019} an efficient method to verify the preparation of Dicke states is proposed. For example, consider a tripartite system and the W state $\ket{D^3_1} = (\ket{001} + \ket{010} + \ket{100})/\sqrt{3}$ as target. The problem here consists in devising an optimal protocol to discriminate between $\sigma = \ketbra{D^3_1}{D^3_1}$ and $\sigma$ such that $\bra{D^3_1}\sigma\ket{D^3_1} = 1-\epsilon$, where 
$\epsilon$ is the so-called infidelity of the imperfect preparation~\cite{Pallinster2018}. The Dicke states do not support a stabilizer $\mathcal{T}$. Nonetheless, such fact does not hinder the construction of optimal verification schemes. Below, we present an efficient adaptive strategy. 
In each run: 
\begin{enumerate}
    \item Choose a party $i\in [3]$ at random and measure $Z$.
    \item If the outcome $+1$ is obtained, measure $Z$ in the two remaining parties; otherwise, if $-1$ is obtained, measure $XX$.  
    \item The test is passed when $Z$ exhibits $+1$ for both parties or $XX$ yields $+1$. In this case, we proceed to the next round; otherwise, the protocol is halted and we conclude that the state $\sigma$ is not $\ketbra{D^3_1}{D^3_1}$.     
\end{enumerate}
For an arbitrary $\sigma$, the probability to pass the test can be expressed as the expectation value of the strategy operator $\Omega_{D^3_1}$:
\begin{equation}
  \Omega_{D^3_1} = (Z^-(Z^+Z^+) + Z^+(XX)^+)_{\rm{PI}}~,
  \label{eq:OmegaD}
\end{equation}
where $A^{\pm} = (\mathds{1}\pm A)/2$ are the projectors onto the $\pm 1$ subspaces of the operator $A$, fulfilling $A^2 = \mathds{1}$. From Eq.~\eqref{eq:OmegaD}, it follows that a perfect preparation $\sigma = \ketbra{D^3_1}{D^3_1}$ always passes the test. On the other hand, if the infidelity of the state is fixed to  $1-\bra{D^3_1}\sigma\ket{D^3_1} = \epsilon$, with $\epsilon$ finite, we need on average $\mathcal{O}(3\epsilon^{-1}\log \delta^{-1})$ shots to observe a test failure with confidence level $\delta$. The previous reasoning can be extended to an arbitrary number of parties to implement a test for the verification of the state $\ket{D_1^N}$. Figure~\ref{fig:DickeVeri} summarises the protocol for this scenario. 

Finally, similar strategies are applicable to generic Dicke states $\ket{D_k^N}$ of arbitrary number of excitations $k$. As a result, only $\mathcal{O}(N\epsilon^{-1}\log \delta^{-1})$ copies suffice to verify $N$-partite Dicke states with an infidelity $\epsilon$ and confidence level $\delta$, which is already exponentially better than traditional tomographic techniques. Moreover, similar protocols have been proposed for phased Dicke, Dicke states of qudits and antisymmetric basis states~\cite{Li2021}. 

The preparation of symmetric states can be certified with stronger tests, which rely on fewer assumptions about the system. Specifically, we adopt the so-called device-independent paradigm, in which parties are modelled as black boxes processing classical information. This framework is particularly useful, as it does not require measurement calibration  nor trust in the internal workings of the setup. 

\subsection{Self-testing}

\noindent Bell inequalities (BIs) allow to certify entanglement from minimal assumptions, namely, in a device independent manner~\cite{brunner_bell_2014}. Their importance was duly acknowledged by the 2022 Nobel Prize in Physics. BIs are also instrumental to self-test quantum states~\cite{Supic2020}. In some instances, the maximal quantum violation of a BI certifies a unique quantum state and measurement implementations up to local isometries. Under these conditions, we conclude that the quantum state is self-tested, representing one of the most stringent verification protocol schemes. This implies that the physical realization of both states and measurements can be certified solely by analysing the statistics of the measurement outcomes. 

In the case of bipartite systems, the simplest example is given by the Clauser-Horne-Shimony-Holt (CHSH) inequality~\cite{Clauser1969}: $\mathcal{B} = \langle A_0B_0 \rangle +  \langle A_0B_1 \rangle + \langle A_1B_0 \rangle - \langle A_1B_1 \rangle\leq 2$, where $\{A_0, A_1, B_0, B_1 \}$ are arbitrary local measurements with binary outcomes and $\langle\cdot \rangle$ represents the expectation value in a purely statistical sense. In particular, for quantum correlations, the expectation values must originate from a proper quantum state $\rho$ via the Born's rule $\langle\cdot \rangle = \mathrm{Tr}(\cdot\rho)$. For such correlations, the achievable maximal violation is $\mathcal{B} = 2\sqrt{2}$~\cite{Cirelson1980}. Remarkably, up to local unitaries, the Bell state $\ket{\Phi^+} = (\ket{00} + \ket{11})/\sqrt{2}$ with appropriate Pauli measurements is the only state which saturates this bound. More generally, for bipartite systems, any pure entangled state can be self-tested~\cite{Coladangelo2017}.   

For its multipartite counterparts, highly entangled states, such as the GHZ and the W state can be robustly self-tested~\cite{Pal2014,Wu2014,Zhou2022}. Such results are generalized for qubit Dicke states $\ket{D^N_k}$ for arbitrary number of parties $N$ and excitations $k$~\cite{Fadel2017, Supic2018}. Recent results establish that any $N$-qubit entangled pure state can be self-tested in a network-scenario~\cite{Supic2023} and in the usual Bell scenario~\cite{balanzo2024all}. However, the number of measurements required for the test generally scales too badly for its experimental implementation.    

A promising scalable relaxation of the problem involves extending the notion of self-testing from states to subspaces, which may define coarse-grained properties of the system. In this regard, \cite{Frerot2021} presents a class of permutation-invariant BIs (PIBIs) which self-tests the zero total spin $J=J_{\min} = 0$ subspace for even $N$ of a PI state (see Eq.~\eqref{eq:SWJ}). The proposed PIBI is based only on two-body correlators and thus it is feasible to probe in quantum many-body setups (e.g. in atomic ensembles~\cite{KAWAGUCHI2012253}). The $J=0$ sector comprises all $N$-partite states that are left invariant under any collective rotation $U^{\otimes N}$, $UU^\dagger = \mathds{I}$. It is massively degenerated and is spanned by all many-body spin singlets, i.e., zero eigenstates of the total spin $\mathbf{J}^2$. According to a theorem in \cite{Auerbach1994}, many-body spin singlets naturally arise as ground states of Heisenberg antiferromagnets~\cite{Arecchi1972} and their realizations in e.g. Fermi-Hubbard models~\cite{Tarruell2018}.

Two-body PIBIs cannot be generically used to self-test states, as the very same two-body marginals may be compatible with different global states. However, some instances of the reduced density matrix uniquely determine the corresponding global symmetric state, as in the case, for instance, of Dicke states~\cite{Aloy2021}. This unique property could open the way to a weaker form of self-testing, under the assumption that the global state is symmetric. In the following, we discuss various approaches to characterise the entanglement content of symmetric states.

\subsection{Entanglement certification}

\noindent The certification of entanglement, that is, deciding whether a given quantum state is entangled or not, can be realized in various ways.
A typical approach relies on the construction of an \textit{entanglement witness} (EW), that is, a Hermitian operator  $W \in \mathcal{H}$ satisfying two properties: i) $\mbox{Tr}(W \rho_{sep}) \geq 0$ for every separable state $\rho_{sep} \in \mathcal{B}(\mathcal{H})$, and ii) there exists at least one entangled state $\sigma \in \mathcal{B}(\mathcal{H})$ such that $\mbox{Tr}(W \sigma) < 0$. From a geometrical point of view, as a consequence of the Hahn-Banach theorem, EWs can be seen as hyperplanes separating the the set of the entangled states they detect from the convex set of separable states. In the context of entanglement detection, particularly relevant is the concept of decomposability of an EW. If $W$ can be expressed as 
\begin{equation}
\label{EW_dec}
    W= P + Q^{T_{B}}~,
\end{equation}
for some positive operators $P$ and $Q$, then $W$ is said to be decomposable; on the contrary, if such decomposition does not exist, it is dubbed non-decomposable. This latter class of entanglement witnesses is extremely interesting, as they are the sole candidates capable of detecting PPT-entangled states (for a comprehensive review of entanglement witnesses we refer the reader to \cite{chruscinski2014entanglement}). 

In the previous section, we have seen that the cone of completely positive matrices, $\mathcal{CP}_{d}$, plays a crucial role in the separability problem for two-qudit DS states. Another interesting feature derives from considering its dual cone, known as the copositive cone, and denoted as $\mathcal{COP}_{d}$. Formally, a $d \times d$ matrix $H$ is said to be copositive if $\vec{x}^{T} H \vec{x} \geq 0$, for every entry-wise nonnegative vector $\vec{x} = (x_{1}, \dots, x_{d-1}) \in \mathbb{R}^{d}$. Similarly to the case of completely positive matrices, also copositive matrices find applications in a variety of problems, mostly related to the field of mathematical optimization (for a review of such problems see \cite{bomze2012copositive}). 
Remarkably, copositive matrices reveal an interesting connection to the concept of entanglement witness. 
In fact, similarly to the case of entanglement witnesses, they are also endowed with a natural concept of decomposition. Indeed, it is a known result that $\mathcal{PSD}_{d} + \mathcal{N}_{d} \subseteq \mathcal{COP}_{d}$, where $\mathcal{PSD}_{d}$ and $\mathcal{N}_{d}$ are the convex cones of positive semidefinite and of entry-wise nonnegative matrices of order $d$, respectively. In particular, it has been proven by Diananda that $\mathcal{COP}_{d} = \mathcal{PSD}_{d} + \mathcal{N}_{d}$ for  $d\leq 4$, while the inclusion is strict for any $d\geq 5$ \cite{diananda_1962}. As a consequence, any copositive matrix $H \in \mathcal{COP}_{d}$ in $d\leq 4$ admits a decomposition of the form 
\begin{equation}
\label{cop_dec}
H = H_{\mathcal{PSD}_{d}} + H_{\mathcal{N}_{d}}~,
\end{equation}
where $H_{\mathcal{PSD}_{d}} \in \mathcal{PSD}_{d}$ and $H_{\mathcal{N}_{d}} \in \mathcal{N}_{d}$. However, when $d \geq 5$, there exist copositive matrices, dubbed \textit{exceptional}, which stand out for their impossibility to be decomposed as in Eq.~\eqref{cop_dec}. 

\begin{figure}[h]
    \centering
    \includegraphics[width=1\linewidth]{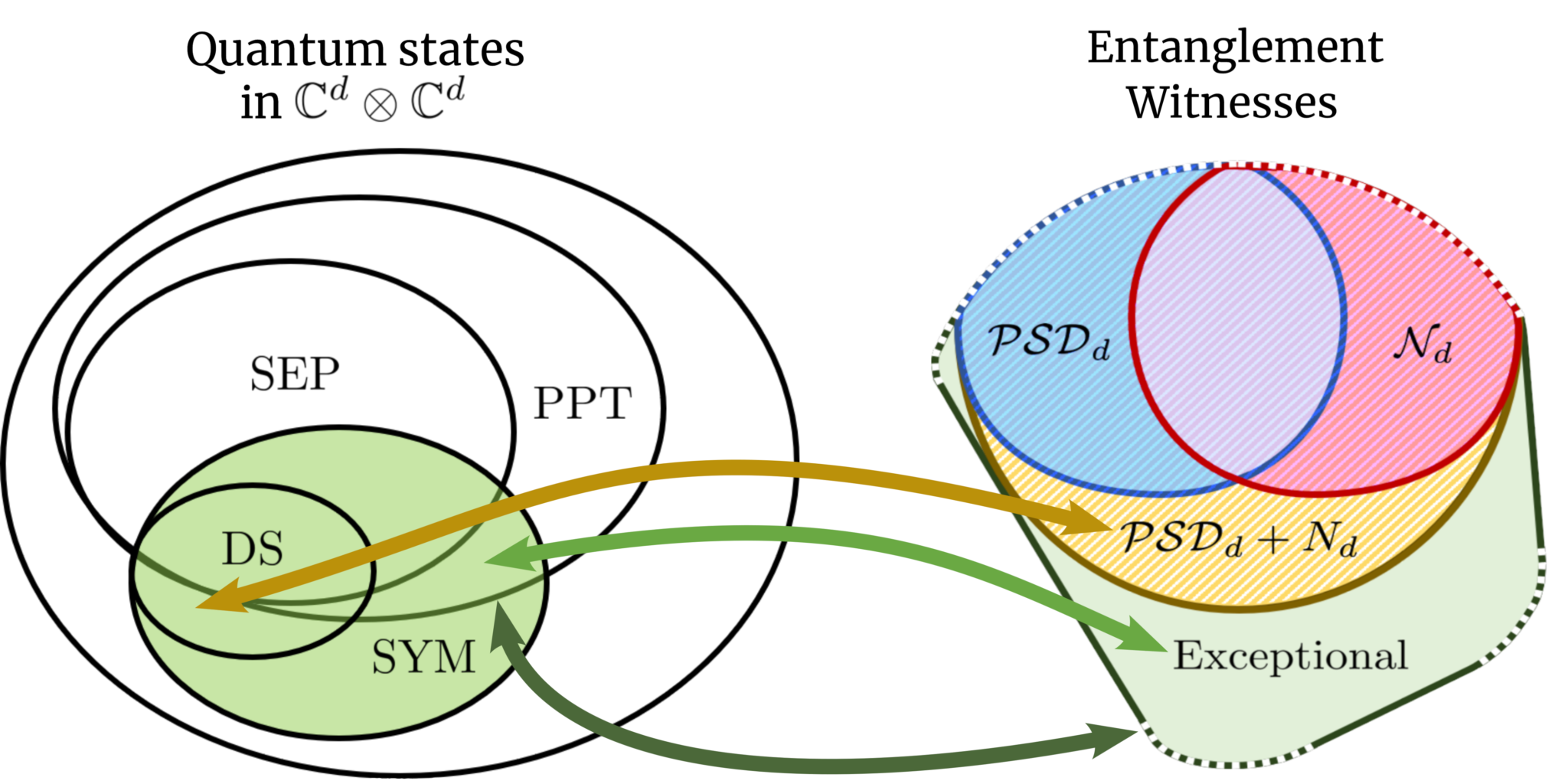}
    \caption{Pictorial representation of the relation between the entanglement properties of two-qudit symmetric states and the cone of copositive matrices acting as entanglement witnesses. The convex cones of positive semidefinite matrices ($\mathcal{PSD}_{d}$) and  nonnegative matrices ($\mathcal{N}_{d}$) are represented in light blue and red, respectively. The yellow region, corresponding to the convex hull of such cones ($\mathcal{PSD}_{d}+ \mathcal{N}_{d}$), has been overmagnified for a clearer visualisation. Such region, corresponding to non-exceptional copositive matrices, is equivalent to the set of decomposable EWs for symmetric states. The green region corresponds to exceptional copositive matrices, which detect PPT-entangled symmetric states and exist only for $d\geq 5$. Extreme copositive matrices (dashed lines) lie at the border of the cone $\mathcal{COP}_{d}$. Notice that not every matrix on the border is extreme.}
    \label{fig:state-ew}
\end{figure}

A closer look to Eqs.~\eqref{EW_dec}-\eqref{cop_dec} shows an intriguing similarity and indeed this analogy has been made formal in \cite{MarconiQuantum2021}. There, it was shown that any copositive matrix, $H=\sum_{i,j=0}^{d-1}H_{ij}\ketbra{i}{j}$, with at least one negative entry leads to an entanglement witness on the symmetric subspace of the form $W=\sum_{i,j=0}^{d-1}H_{ij}\ketbra{ij}{ji}$. 
Moreover, it was also proved that there exists a one to one correspondence between exceptional copositive matrices, such that $H \neq H_{\mathcal{PSD}_{d}} + H_{\mathcal{N}_{d}}$, and non-decomposable entanglement witnesses $W \neq P + Q^{T_{B}}$ for symmetric states.
This correspondence can be pushed further introducing the concept of an extreme copositive matrix $H_{ext}$ \cite{hiriart2010variational}, for which $H_{ext} = H_1 + H_2$, with $H_1 , H_2 \in \mathcal{COP}_{d}$, implies $H_1 = a H_{ext}, H_2 = (1 - a) H_{ext}$ for all $a \in [0, 1]$. 
Indeed, in \cite{marconi2023entanglement} it was proved that extreme copositive matrices give rise to optimal entanglement witnesses for bipartite symmetric qudits, which are tangent to the set of separable states. A schematic overview of these concepts and their relationships is presented in Figure \ref{fig:state-ew}.

In the case of many-body systems, EWs based on spin-squeezed inequalities can be used as a tool for entanglement certification. This approach is particularly suited to the case of many-body spin systems, for which the global spin operators can be defined as as $J_\alpha=\sum_{i=1}^N \frac{1}{2}\sigma_{i}^{\alpha}$, with $\alpha=\{x,y,z\}$. Hence, a state with total spin $J$ is said to be \textit{spin-squeezed} \cite{kitagawa1993squeezed} if 
\begin{equation}
    \xi^{2} \equiv \frac{(\Delta J_{\bm{n}_1})^2}{\langle J_{\bm{n}_{2}}\rangle^2 +\langle J_{\bm{n}_{3}}\rangle^2 } < 1~,  
\end{equation}
where $(\Delta J_{\bm{n}})^2 \equiv \langle J_{\bm{n}}^2\rangle  - \langle J_{\bm{n}}\rangle^2 $ and $J_{\bm{n}}\equiv \bm{n} \cdot \bm{J}$, for a given direction $\bm{n}=(\bm{n}_{1}, \bm{n}_{2}, \bm{n}_{3})$. 
Remarkably, in \cite{sorensen_many-particle_2001}, a criterion to detect entanglement in spin-squeezed states was found. Expanding on this result, the relation between spin-squeezing inequalities and entanglement in many-body systems was explored in many other works (see \cite{vitagliano2014spin} and references therein). In \cite{PhysRevLett.95.120502,PhysRevLett.95.259901,Vidal_2006}, some inequalities were constructed for symmetric $N$-qubit states, providing necessary and sufficient criteria to certify two- and three-qubit genuine multipartite entanglement, while being only necessary for a higher number of qubits. Nevertheless, other approaches to detect GME around qubit Dicke states have also been developed \cite{t_th_2007}, showing that in the vicinity of Dicke states there always exists GME.

\section{Applications of symmetric quantum states}
\label{sec:Applications}
\noindent  Due to their entanglement properties and robustness against noise, symmetric states are highly appealing to various emergent quantum-enhanced technologies, such as sensing, computation and simulation. 
We begin by reviewing their applications in quantum  metrology with emphasis in the improvements that arise from using different classes of symmetric or PI states such as the GHZ or squeezed states. Next, we focus on the use of symmetric states in quantum error correction and conclude by addressing their importance in quantum communication and computing.  

\subsection{Quantum metrology}
\noindent The formalism of quantum mechanics predicts the existence of incompatible observables, which do not have a well-defined joint value within the same context \cite{budroni2022kochen}. Such counterintuitive fact explains, for instance, the zero-point fluctuations observed in isolated quantum systems. These fluctuations are fundamental in nature and arise as an unavoidable noise in highly demanding experimental situations, even when all other technical sources of instability, such as vibrations, temperature fluctuations, and other disturbances, have been efficiently suppressed. Driven by both fundamental curiosity and practical interest, it is quite remarkable that we have managed to tame this uncertainty to unexpectedly improve measurement precision~\cite{Frowis2015}. Here, we refer to measurements of physical quantities, such as magnetic fields, distances, time intervals, which are not associated to any quantum-mechanical observable. Quantum metrology aims at increasing the precision of such measurements by exploiting quantum properties such as entanglement. Precision measurements are very often obtained by mapping the physical quantity of interest to a phase shift that can be determined using interferometric techniques~\cite{Pezze2018}. Thus, interferometry stands out as a unifying framework to study precision measurements via a phase estimation task.\\ 

\noindent \textbf{The phase estimation task}. The phase estimation task involves a probe state, $\rho$, in which a phase $\theta$ is encoded via a unitary transformation, that models an interferometer. The output state is given by $\rho(\theta) = e^{-i\theta G}\rho e^{+i\theta G}$, where $G$ is the generator of the transformation. Then, the goal is to infer the parameter $\theta$ with the maximal precision by measuring an observable $O$ on the transformed state $\rho(\theta)$. According to the error propagation formula, the precision of such task, as quantified by the inferred variance $(\Delta \theta)_{\mathrm{est.}}^2$, is bounded by the signal-to noise-ratio as follows: 
\begin{equation}
\label{eq:sens_bound}
\nu^2(\Delta\theta)^2_{\mathrm{est.}}\geq \frac{(\Delta O)^2}{\langle i[G, O] \rangle^2 }:=\frac{1}{\Xi^{-2}_O[\rho, G]},
\end{equation}
where $\nu\gg 1$ is the number of experimental runs or shots which we assume to be large, $(\Delta O)^2 = \langle O^2 \rangle-\langle O \rangle^2$ is the variance, and $\langle O\rangle = \mathrm{Tr}(O\rho)$ is the mean value of the observable $O$.
For generic encoding routines, $\rho(\theta) = \Lambda_\theta(\rho)$, beyond the case of unitary transformations, one simply has to replace $\langle i[G, O] \rangle $ by $\partial_\theta\langle O\rangle_{\rho(\theta)}$. The parameter $\Xi^{-2}_O$ of Eq.~\eqref{eq:sens_bound} represents the sensitivity of $O$ for the estimation task. The best sensitivity
\begin{equation}
\label{eq:QFI}
    \max_{O=O^\dagger} \Xi_{O}^{-2}[\rho, G] = F_Q[\rho, G],
\end{equation}
corresponds to the so-called quantum Fisher information (QFI) and constitutes the ultimate precision bound for the phase estimation task~\cite{Braunstein1994}. 

By means of the Heisenberg-Robertson inequality $(\Delta G)^2(\Delta O)^2\geq \langle i[G,O] \rangle^2/4 $, we can establish an upper bound for the QFI, i.e., $F_Q[\rho, G]\leq 4(\Delta G)^2$. Interestingly, this bound is saturated by pure states $\rho = \ketbra{\Psi}{\Psi}$. Hereinafter, we will partition the system into $N$ qubits (or spin-$1/2$ particles) and consider collective encoding, i.e., generators of the form $G = \sum_{i=1}^NZ_i/2 = J_z$. As such, the encoding operation corresponds now a collective rotation around an axis, which is often fixed to $z$ by convention. This class of transformations are routinely implemented in Ramsey interferometry setups by letting the system evolve under strong homogeneous magnetic fields or microwave pulses~\cite{Ramsey1949}. Taking advantage of the previous observation, one can compute the maximal QFI extractable from a $K$-producible state (see Section \ref{sec:entanglement} for the definitions of $K$-producibility and entanglement depth). By doing so, the following bound is established~\cite{Hyllus2012, Toth2012}:  
\begin{equation}
 \label{eq:FQKN}
    F_Q[\rho_K,J_z]\leq KN,
\end{equation}
where $\rho_K$ is any state which is $K$-producible. Tighter bounds were established in the same references. Nonetheless, Eq.~\eqref{eq:FQKN} is of fundamental importance since it implies that, in order to improve the sensitivity of an experiment, it is mandatory to either increase its size $N$ or its entanglement content $K$. The separable bound $K = 1$ is usually referred to as the shot-noise or standard quantum limit (SQL), whereas the quantum bound $K = N$, which is necessarily achieved by genuinely entangled states, is called the Heisenberg limit. Similar bounds exist for other scalable quantifiers of multipartite entanglement such as $L$-separability~\cite{PhysRevA.91.042313}, entanglement
depth of formation, the squareability entanglement depth and its combined
convex roofs~\cite{Szalay2025alternativesof}. Hence, the QFI is a versatile multipartite entanglement quantifier and it is employed to witness metrologically-useful entanglement in many-body states. In fact, metrology constitutes one of the few confirmed quantum-enhanced applications in which multipartite entanglement acts manifestly a resource.

\subsubsection{Symmetric states are maximally metrologically useful}

\noindent Symmetric and permutation-invariant states play a central role in metrological applications. Remarkably, random multipartite pure states drawn from the symmetric subspace typically exhibit Heisenberg scaling, in stark contrast to generic pure states, which typically do not~\cite{OszmaniecPRX2016}. 
In the following paragraphs, we discuss the metrological properties of the most commonly used symmetric and PI states. \\

\noindent\textbf{GHZ states}.-- Let us consider states with total spin length $ \sqrt{J(J+1)}$. Then, the variance of $J_z$ cannot surpass $J^2$. Hence, $F_Q[\rho, J_z]\leq 4(\Delta J_z)^2\leq 4J^2$, and we reach the conclusion that the most sensitive states have maximal spin  $J = J_{\max} = N/2$, and are therefore supported in the fully symmetric sector. The canonical example of such states is the GHZ state, also denoted NOON in the context of bosonic particles:
\begin{equation}
\ket{\rm{GHZ}} = \frac{1}{\sqrt{2}}(\ket{0}^{\otimes N} + \ket{1}^{\otimes N}) \ .
\label{eq:GHZ}
\end{equation}
The GHZ is unpolarised, i.e., $\bra{\rm{GHZ}}J_x\ket{GHZ} = \bra{\rm{GHZ}}J_y\ket{GHZ} = \bra{\rm{GHZ}}J_z\ket{GHZ} = 0$, and indeed achieves the Heisenberg limit; $\bra{\rm{GHZ}} J_z^2\ket{\rm{GHZ}} = (N/2)^2$. The optimal observable $O$ saturating the QFI is the parity operator along $x$, $(-1)^{J_x+N/2}:=\Pi_x$, or, in fact, any direction orthogonal to $z$. It is also instructive to verify its metrological usefulness directly from the interferometric protocol. After acquiring the phase, the state is transformed to $e^{-i\theta J_z}\ket{\rm{GHZ}} = (e^{iN\theta/2 }\ket{0}^{\otimes N} +  e^{-iN\theta/2}\ket{1}^{\otimes N})/\sqrt{2} = \ket{\rm{GHZ}(\theta)}$. Its expectation value and variance against the parity operator are  $\langle \Pi_x \rangle(\theta) = \cos(N\theta)$ and $(\Delta \Pi_x)^2(\theta) = \sin^2(N\theta)$ respectively. Hence, the sensitivity exhibited by this state is maximal $\Xi^{-2}_{\Pi_x} = [\partial_\theta \langle\Pi_x \rangle(\theta)]^2/(\Delta \Pi_x)^2(\theta) = N^2$. \\

\noindent\textbf{Singlet states}.-- In the opposed extreme is found $N$-partite singlet sector, $J=0$. As already announced, this manifold contains only states $\rho$ that are fully rotation-invariant, $[J_x,\rho] = [J_y, \rho] = [J_z, \rho] = 0$, and generalize the ordinary spin singlet $(\ket{01}-\ket{10})/\sqrt{2}$ to arbitrary even $N$. Due to this symmetry, the singlet is insensitive to the phase imprinting protocol introduced in this section. As a result, it is useless for the estimation of rotation angles. However, they are specially suited to infer instead gradients of magnetic fields~\cite{Urizar2013}. \\

\noindent\textbf{Dicke states}.-- Let us now move to Dicke states. In particular, consider the balanced Dicke $\ket{D_{N/2}^N}$, which in the spin language corresponds to $\ket{J= N/2, M = 0}$. It exhibits no polarization and is an eigenstate of $J_z$. Hence, it remains invariant under rotations around this axis. However, under rotations orthogonal to $z$, it displays exquisite sensitivity $(\Delta {J}_x)^2 = (\Delta {J}_y)^2 =[(N/2)(N/2 +1)]/2 $. As reviewed in the following section, Dicke states are realized in various experiments. Here, we want to highlight the preparation of $N =10000 $ Dicke in a spin-1 Bose-Einstein condensate~\cite{Zou2018} and its subsequent demonstration of beyond shot-noise interferometric sensitivity. \\

\noindent \textbf{Coherent spin states}.-- Next, we inspect polarised states, starting with those with maximal mean spin $\langle J_{x} \rangle^2 + \langle J_{y}\rangle^2 + \langle J_{z}\rangle^2 = {J}_{\max}^2$, with $J_{\max} = N/2$. They can be parametrised from a unit vector in $\mathbf{v}\in \mathbb{R}^3$ as the extremal eigenstate of the collective spin projection, $J_{\mathbf{v}} = \mathbf{v}\cdot\mathbf{J}$, where $\mathbf{J} = (J_x, J_y, J_z)$. Provided $\mathbf{v}$, such state is unique and corresponds to the so-called coherent spin state (CSS), $\ket{\mathrm{CSS}_{\mathbf{v}}} = \ket{+1_{\mathbf{v}}}^{\otimes N}$, with $\ket{+1_{\mathbf{v}}}$ being the $+1$ eigenstate of $\mathbf{v}\cdot\mathbf{j}$ with $\mathbf{j} = (X,Y,Z)$. Notice that the CSS is a product (separable) state. Along the orientation $y = (0,1,0)$, it reads (cf. the GHZ state Eq.~\eqref{eq:GHZ}): 
\begin{equation}
  \ket{\mathrm{CSS}_y} =   \left[\frac{1}{\sqrt{2}}(\ket{0} + i\ket{1})  \right]^{\otimes N}.
  \label{eq:CSSx}
\end{equation}
Now, we proceed to assess the metrological usefulness of the CSS Eq.~\eqref{eq:CSSx}. To this end, we consider rotations around $J_z = G$ and set $J_x = O$ as probe observable. Then, we evaluate $\langle i [J_z, J_x] \rangle = \langle J_y \rangle = J_{\max} = N/2$ and $(\Delta J_x )^2 = N/4$, leading to a sensitivity:
\begin{equation}
    \Xi^{-2}_{J_x} = \frac{\langle J_y \rangle^2}{(\Delta J_x)^2} = N ,
\label{eq:XiCSS}
\end{equation}
which attains the QFI $F_Q[\ketbra{\mathrm{CSS}_y}{\mathrm{CSS}_y}, J_z]=4(\Delta J_z )^2 = N$ and, more importantly, saturates the shot-noise bound Eq.~\eqref{eq:FQKN}. Hence, we deduce that the CSS is the maximally metrologically useful separable state and consequentially, that entanglement is needed to improve the sensitivity Eq.~\eqref{eq:XiCSS}. \\

\noindent\textbf{Spin-squeezed states.--} How can one enhance the CSS's sensitivity? To do so, it is possible to reduce variance $(\Delta J_x)^2$ while preserving a large value of the mean spin $\langle J_y \rangle$ via spin-squeezing protocols. As a result, the sensitivity $\Xi^{-2}_{J_x}$, is increased. The variance in $z$ and $x$ are not independent and are constrained by the uncertainty relation, $(\Delta J_z)^2(\Delta J_x)^2 \geq \langle J_y \rangle^2/4$. The CSS saturates this relation by presenting the variances $(\Delta J_x)^2 = (\Delta J_y)^2 =  \langle J_y \rangle/2 = N/4 $. Then, the squeezing strategy boils down to redistribute the noise and suppress enough the fluctuations in the observable $J_x$, $(\Delta J_x )^2$. Such fact implies an increase of $(\Delta J_z)^2$, and with it, an improvement of the QFI, as well. States possessing large mean spin and minimal fluctuations orthogonal to it are termed spin-squeezed states (SSS)~\cite{Kiagawa1993}. The amount of squeezing is quantified by the sensitivity normalized to its coherent value, $\Xi^{-2}_{J_x}/N:=\xi^{-2}_{x}$, which within this framework, it is also indicated as metrological spin-squeezing parameter. In the laboratory, SSSs can be generated by entangling CSS via nonlinear spin-mixing evolution. For instance, via the celebrated one-axis twisting (OAT) dynamics, which evolves the initial CSS as per  
\begin{equation}
    \ket{\Psi(t)} = e^{-itJ_x^2}\ket{\mathrm{CSS}_y} \ .
\end{equation}
At short preparation times $t\sim\mathcal{O}(N^{-2/3}) $, the state $\ket{\Psi(t)}$ is maximally spin-squeezed delivering sensitivities up to $\xi^{-2}_{J_a}\sim \mathcal{O}(N^{2/3})$ for the duly calibrated probe.

\begin{figure}
    \centering
    \includegraphics[width = 0.6\columnwidth]{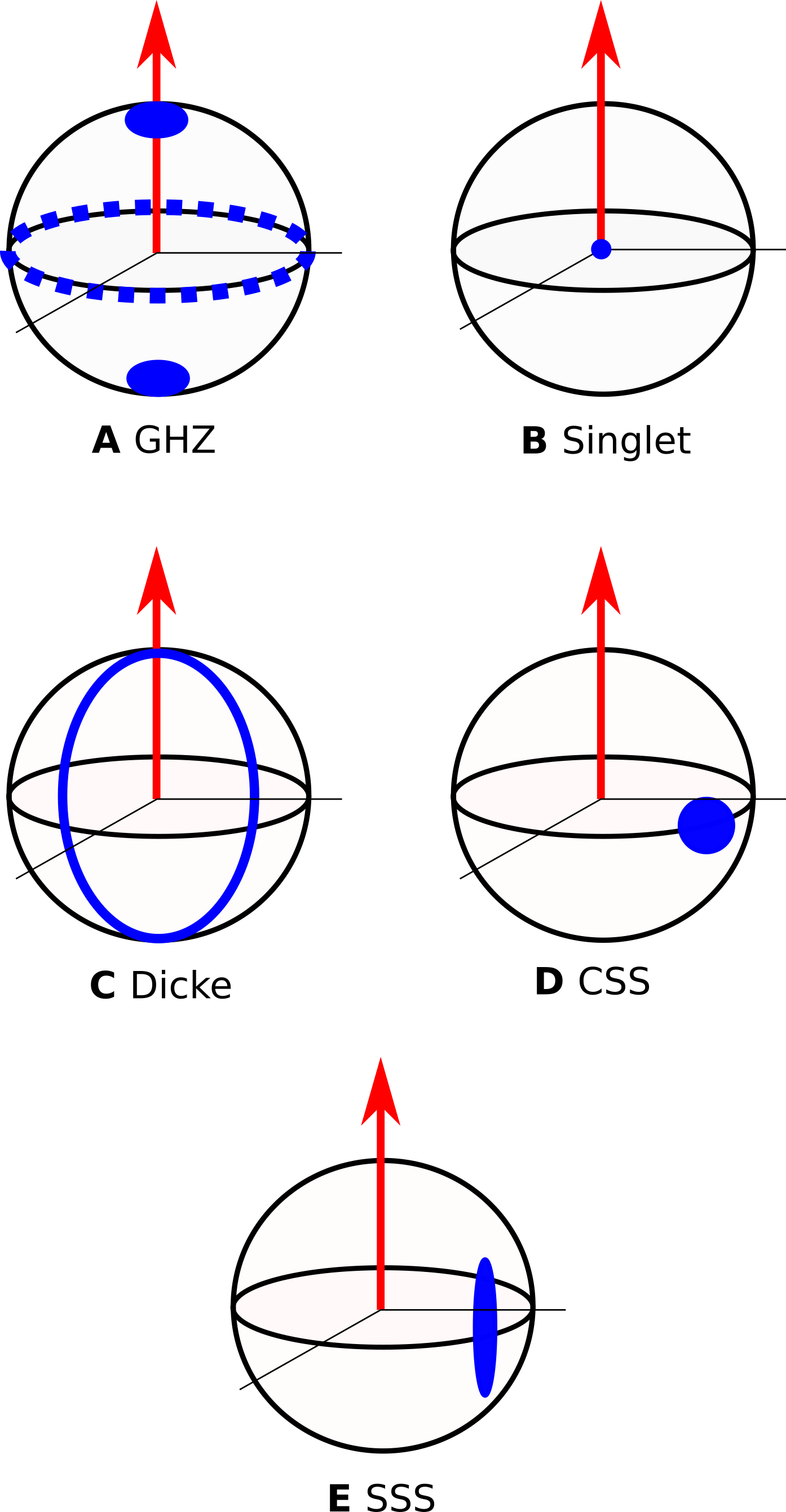}
    \caption{Phase space representation in the collective Bloch ball of the many-body spin states described in this subsection. The red arrow represents the axis with respect to which they exhibit maximal sensitivity. Source: original.}
    \label{fig:Fig_metro} 
\end{figure}

As previously discussed, the totally symmetric space of $N$-qubits (or spin-$1/2$) can be mapped, up to permutation symmetry, to a spin representation resulting from the coherent participation of the $N$ elementary spins. In this collective spin picture, such states may be represented in a ``giant'' Bloch ball of radius $J_{\max} = N/2$ via Husimi quasiprobabilities or other phase-space methods. This visualization is very useful to asses at a glance the metrological usefulness of the state under consideration. In Figure~\ref{fig:Fig_metro}, we sketch the phase-space picture associated to the classes of states addressed above. 

The GHZ state (panel A) can be represented as a combination of two spin coherent states in the opposite poles (blue blobs) plus a series of fringes around the equator stemming from the interference of both coherent components: $2\ketbra{\mathrm{GHZ}}{\mathrm{GHZ}} = \ketbra{\rm{CSS}_{z}}{\rm{CSS}_{z}} + \ketbra{\rm{CSS}_{-z}}{\rm{CSS}_{-z}} + \ketbra{\rm{CSS}_{z}}{\rm{CSS}_{-z}} +\ketbra{\rm{CSS}_{-z}}{\rm{CSS}_{z}} $.It is precisely these coherences that endow the GHZ state with its exquisite sensitivity. The singlet state (panel B), fulfilling $(\Delta J_x)^2 = (\Delta J_y)^2 = (\Delta J_z)^2 = 0 $, is depicted as a point in the centre of the sphere, which indeed remains unchanged under any rotation. The balanced Dicke state (panel C) is represented as a ring on the equator of the sphere, which is invariant under rotations around one axis $a$, being an eigenstate of $J_a$. The coherent spin state (panel D) corresponds to an isotropic blob in a certain direction described by the vector $\mathbf{v}$. Finally, one can deform the blob to enhance its sensitivity as in the case of the spin-squeezed state (panel D).

\subsubsection{Spin squeezing criteria: Bell correlation, entanglement and metrological usefulness}

\noindent We conclude this subsection by further exploring the connection between metrological usefulness and entanglement in the PI sector. Note that by combining the sensitivity of spin measurements Eq~\eqref{eq:XiCSS} and the separable bound for QFI Eq.~\eqref{eq:FQKN}, the metrological spin-squeezing parameter defines an entanglement witness~\cite{Wineland1994, Sorensen2001}:
\begin{equation}
\frac{\langle J_y \rangle^2_{\rho}}{(\Delta J_x)^2_{\rho}} < N \Longrightarrow \mbox{$\rho$ entangled} \ .
\label{eq:SSIneq}
\end{equation}
Note that only low order moments of the collective spin (up to quadratic) are needed to evaluate the witness. Such data can be readily be inferred in Bose-Einstein condensates experiments. Extensions for higher orders for the spin squeezing inequality~\eqref{eq:SSIneq} are also established~\cite{Gessner2019}. Furthermore, in \cite{MullerRigat2023}, the authors report a data-driven method to bound directly the QFI from a given set of few mean values. The approach enables to recover stronger versions of spin-squeezing inequalities. The most recent advances suggests that symmetry and coherence are a resources for metrology~\cite{Frerot2004}. In \cite{fadel_bounding_2020}, the authors present an algorithm to bound the fidelity from permutation-invariant observables such as the low-order moments of the collective spin that we considered in this section.  

As already noted, the witness Eq.~\eqref{eq:SSIneq} contains only low order moments of PI observable. One may therefore ask if such symmetries are also useful to simplify the derivation of generic entanglement witness, which detect states which are not necessarily metrologically useful,e.g., the singlet~\cite{Toth2009}. In this regard, \cite{Toth2009b} remarkably provides a complete set of only $8$ inequalities that are sufficient to capture all entanglement that can be certified from the mean spin and its fluctuations: 
\begin{align}
    \label{Toth}
   &\sum_{a\in\{x,y,z\}}  \langle J_a^2\rangle \leq N(N+2)/4 \\
    &  \sum_{a\in\{x,y,z\}}  ( \Delta J_a)^2 \geq N/2 \\
    &\langle J_a^2 \rangle +  \langle J_b^2 \rangle -(N-1)(\Delta J_c)^2 \leq N/2\\
    &(N-1)((\Delta J_a)^2 +(\Delta J_b)^2)-\langle J_c^2\rangle\geq N(N-2)/4,
    \end{align}
where $\{a,b,c \}$ vary over all permutations of $\{x,y,z \}$. Similar results were generalized to arbitrary spin~\cite{Vitagliano2011}, however, for ensembles of qubits of dimension $d>2$, it does not exist a complete set of inequalities anymore. In \cite{MullerRigat2022}, the authors propose an algorithm to derive the best witness to certify entanglement from second moments of arbitrary observables, beyond spin and illustrate the method by unveiling new experimentally-accessible entanglement criteria based on population measurements onto hyperfine sublevels.  \\

\noindent \textbf{Bell correlation}.-- Interestingly, spin-squeezed states also violate maximally a Bell inequality -- in particular the PIBI presented in \cite{tura2014detecting} and reviewed in Section~\ref{sec:Correlations}. Such PIBI is based on two dichotomic measurements by party $\{m_0, m_1 \}$. In particular, for $\{m_0 = \cos\theta X + \sin\theta Y, m_1 = \cos\theta X - \sin\theta Y \}$, one can express the Bell operator as a function of the angle $\theta$ and the collective operators $J_x^2, J_y$. Then, one may optimize $\theta$ to maximize the changes to reveal nonlocality from a fixed values of $\langle J_x^2\rangle, \langle J_y\rangle$. In so doing, the following Bell correlation witness is derived,
\begin{equation}
\label{eq:SSSBell}
\frac{4\langle J _x^2 \rangle}{N}\geq \frac{1}{2}\left[1-\sqrt{1-\left(\frac{2\langle J_y\rangle}{N}\right)^2}\right],
\end{equation}
whose violation of them signals Bell correlation or Bell nonlocality if the statistics appearing in the equation are inferred via the standard Bell scenario. We use the term Bell nonlocality when we infer the necessary statistics via e.g. collective measurements as expressed in Eq.~\eqref{eq:SSSBell}. In so doing, one needs to assume the validity of Quantum Mechanics and fine calibration of the measurement settings, prompting a departure from the device-independent scenario.  

We observe that spin squeezed states, i.e. states with large mean spin $\langle J_y \rangle$ and small fluctuations $\langle J _x^2\rangle$, violate Eq.~\eqref{eq:SSSBell} the most. However, it is noted that beating the Bell witness Eq.~\eqref{eq:SSSBell} is much more demanding that certifying entanglement via the spin-squeezing inequality Eq.~\eqref{eq:SSIneq} -- states that are not squeezed enough will not be detected by the present criteria. Such observation agrees with the fact that Bell nonlocality is a more stringent resource as compared to its device-dependent counterparts as quantum entanglement. By employing PIBIs with more settings, one can find less demanding Bell witnesses certifying spin-squeezed states~\cite{MullerRigat2021}. Likewise, in \cite{MullerRigat2024}, the authors contribute to new Bell correlation witnesses tailored to spin-nematic squeezing. This type of squeezing manifests in multilevel many-body systems, such as spin-1 BECs~\cite{Hamley2012}.  

The connection between Bell nonlocality and metrological usefulness from these symmetrized correlators is explored further in \cite{Frowis2019} and \cite{Niezgoda2021}. In this last Reference, from BIs discovered in earlier works~\cite{Niezgoda2020,Chwedenczuk2022}, the authors discuss the role of Bell nonlocality itself as a resource for metrology by devising new bounds. 

\subsection{Quantum computing}
\noindent Next, we review the most prominent applications of symmetric states in quantum computing, starting from their role in quantum error correcting codes. 

\subsubsection{Quantum error correction}
\noindent It is an elementary result in quantum information that unknown states cannot be perfectly cloned. Thus, as opposed to classical information, quantum data cannot be made tolerant to errors by simply exploiting the redundancy of multiple copies. Quantum error correction aims at mitigating and protecting quantum devices from imperfections and noise~\cite{Devitt2013}. A plethora of different strategies have been introduced since its first proposal~\cite{Shor1995}, including the harnessing of symmetric states~\cite{Pollatsek2003,Ouyang2014, Albert2022}. Below we review an example of them -- the so-called \textit{gnu code}.  \\

\noindent\textbf{Permutation-invariant codes.}-- The main principle of quantum error correction is to encode the logical bit in multipartite entangled states, which are insensitive to certain errors. The most common approach to accomplish this is to require that the logical qubit is protected by a certain symmetry, e.g. through a set of stabilizers, preventing symmetry-breaking errors. The gnu PI code that we outline here is not based on the stabilizer formalism, and rather encodes the logical bit $\{0_L,1_L\}$ in a superposition of Dicke states as follows
\begin{align}
  &\ket{0_L} = \sum_{\substack{l\mbox{ even} \\ l\in [n+1]}}\sqrt{\frac{{n\choose l}}{2^{n-1}}}\ket{D_{gl}^{gnu}}~, \\
&\ket{1_L} = \sum_{\substack{l\mbox{ odd}\\ l\in [n+1]}}\sqrt{\frac{{n\choose l}}{2^{n-1}}}\ket{D_{gl}^{gnu}}~.
\label{eq:gnu}
\end{align}
The encoding depends on three parameters $\{g,n,u \}$ such that their product amounts to the total number of qubits $gnu = N$. The logical X and Z operations are implemented with the gates $X_L = X^{\otimes N}, Z_L =(e^{i\pi Z/(2g)})^{\otimes N}$. The code gap $g$ corresponds to the spacing between the populated excitation numbers, with a mean value of $gn$, where $n$ where is referred to as the code occupancy. In the limit of large $n$ the binomial weights Eq.~\eqref{eq:gnu} become Gaussian. In such a limit, the approach illustrated here becomes equivalent to the Gottesman-Kitaev-Preskill (GKP) code derived from squeezed harmonic oscillator modes~\cite{Gottesman2001}. The parameters can be adjusted to yield different error correction features. For instance, if $g = t+1, n>3t, u\geq 1 +t/(gn)$, the gnu code corrects $t$ spontaneous decay errors. Whereas, if $g=n=2t+1, u\geq 1$, it is able to correct arbitrary $t$-qubit errors. Let us see how for the simplest case $t=1$. \\

For $t=1$, the gnu code is equivalent to the 9-qubit Ruskai code~\cite{Ruskai2000}, with encoding: 
\begin{align}
  &\ket{0_L} = \frac{\ket{D_0^9} + \sqrt{3}\ket{D^9_6}}{\sqrt{4}}~, \\
    &\ket{1_L} = \frac{\sqrt{3}\ket{D_3^9} + \ket{D^9_9}}{\sqrt{4}}. 
\label{eq:ruskai}
\end{align}
Let us consider the action of Pauli operators $\{X, Y, Z \}$ at party $i = 0$ which perturbs the logical encoding $\{\ket{0_L}, \ket{1_L}\}$. In order to determine how the errors affect the system, we study the Gram matrix: 
\begin{equation}
    M_{sA,rB} = \bra{s_L}A^\dagger B\ket{r_L},
\end{equation}
where $s, r\in {0,1}$, $A,B\in \{X,Y,Z\}$. Direct calculation shows how $M$ factorizes as per $ M_{sA,rB} \propto\delta_{sr}\delta_{AB}$, where $\delta$ is the Kronecker symbol. The structure of the matrix $M$ tells us that it is possible to determine which type of error has occurred ($X$, $Y$ or $Z$) without destroying the coherence of the logical state and restore it by means of unitary transformations. The previous observation stems from the fact that the errors send $\{\ket{0_L}, \ket{1_L} \}$ to two-dimensional orthogonal subspaces corresponding to each type of error~\cite{Laflamme1996}. Checking how the system fixes elementary errors $\{X,Y,Z\}$ is sufficient for correctability of an arbitrary linear combination of them (i.e. a flip in a tilted Bloch orientation) or a statistical mixture of them (e.g. implemented by a Pauli channel). Moreover, by PI invariance, the code corrects equally well a flip in a different location $i\neq 1$, even if such location remains unknown. More generally,$gnu$ codes can correct any number of Pauli errors as well as spontaneous decay and deletion errors \cite{Ouyang2021}.
Such class of codes has been extended to encode multiple  qubit systems~\cite{Ouyang2016}.
Recently, a generalization of several known PI codes has been proposed, where fewer qubits are needed ~\cite{Aydin2024}.
Finally, an approach based on certain polynomials has allowed to construct families of PI codes also for qudit systems \cite{ouyang2017permutation}.

\subsubsection{Quantum algorithms}

\noindent We conclude this subsection by observing that Dicke states find applications also in certain quantum algorithms. For instance, in \cite{d2004computational} it was shown that GHZ and W states are the only pure multipartite states able to solve exactly some computational problems over a complex network. Further, Dicke states have recently emerged as valuable resources for certain variational algorithms, such as the quantum approximate optimization
algorithms (QAOAs), designed to solve classically unfeasible optimization problems \cite{he2023alignment,Mansy2023,Blekos2024}.
Additionally, the framework of permutation-invariant quantum circuits proposed \cite{Mansy2023} further reinforces the computational significance of symmetric states like Dicke states. Their work demonstrates that exploiting permutation symmetry in circuit design can lead to more efficient representations and potentially lower resource requirements for quantum algorithms.

\subsection{Quantum communication and entanglement distribution}

\noindent Symmetric multipartite entangled states, such as Dicke, W, and GHZ states, have proven to be essential resources for a wide range of quantum communication tasks. Their robust entanglement properties make them particularly well-suited for scenarios involving multiple users and distribution across complex networks.
In \cite{roga2023efficient}, a protocol is proposed for efficiently distributing symmetric Dicke states over a lossy quantum network using beam-splitter interference and heralded detection. The authors show that this method enables the scalable generation of multipartite entanglement even in the presence of transmission loss, offering a practical route for deploying Dicke states across quantum networks.
The distribution of multipartite entanglement over noisy networks was investigated also in \cite{bugalho2023distributing}, where the authors presented an algorithm for GHZ states.
On a different note, W states have found applications in protocols requiring robustness against particle loss. In \cite{lipinska2018anonymous}, a protocol for anonymous quantum transmission is introduced, whereby a quantum message can be sent between two parties in a network without revealing their identities. This is achieved using a shared three-qubit W state, which allows to maintain anonymity and security even in noisy environments.
Along a similar line of research, in \cite{miguel2023quantum} a quantum repeater architecture is designed to distribute three-qubit W states over arbitrary distances in a two-dimensional triangular network. The proposed protocol combines entanglement swapping and purification strategies tailored to W states.
Deterministic protocols for entanglement distribution using four-photon Dicke states were also presented in \cite{wang2016deterministic}. The experimental demonstration of other quantum networking protocols, such as quantum telecloning and open-destination teleportation, is reported in \cite{chiuri2012experimental}.

In the context of quantum key distribution (QKD), \cite{zhu2015w} introduces a multi-party measurement-device-independent QKD (MDI-QKD) scheme using four-qubit W states. By designing a W-state analyzer based on linear optics, the authors demonstrate how such entangled states can serve as the basis for secure key exchange among multiple parties, eliminating vulnerabilities associated with measurement devices.
GHZ states, on the other hand, are particularly valuable in scenarios requiring perfect multipartite correlations. The early experimental validation of GHZ entanglement swapping is reported in \cite{lu2009experimental}, where a three-photon GHZ state is generated from independent entangled pairs through projection onto a GHZ basis. This work demonstrates the feasibility of multi-party entanglement distribution and is foundational for applications such as quantum telephone exchange and distributed quantum computation. \cite{li2006efficient} proposed a symmetric quantum state sharing scheme where an arbitrary $N$-qubit quantum state is distributed among multiple parties using $N$ GHZ states. The protocol requires only Bell-basis and single-qubit measurements, achieving high efficiency and security with minimal quantum resources.
In \cite{pickston2023conference}, a GHZ-based protocol for conference key agreement is experimentally demonstrated using a six-user photonic quantum network. The protocol allows all parties to share a secure key directly from a single GHZ state, offering a scalable alternative to conventional pairwise QKD schemes.

\section{Symmetric states in the lab}
\label{sec:Implementations}

\begin{figure}[h!]
    \centering
    \includegraphics[width = \columnwidth]{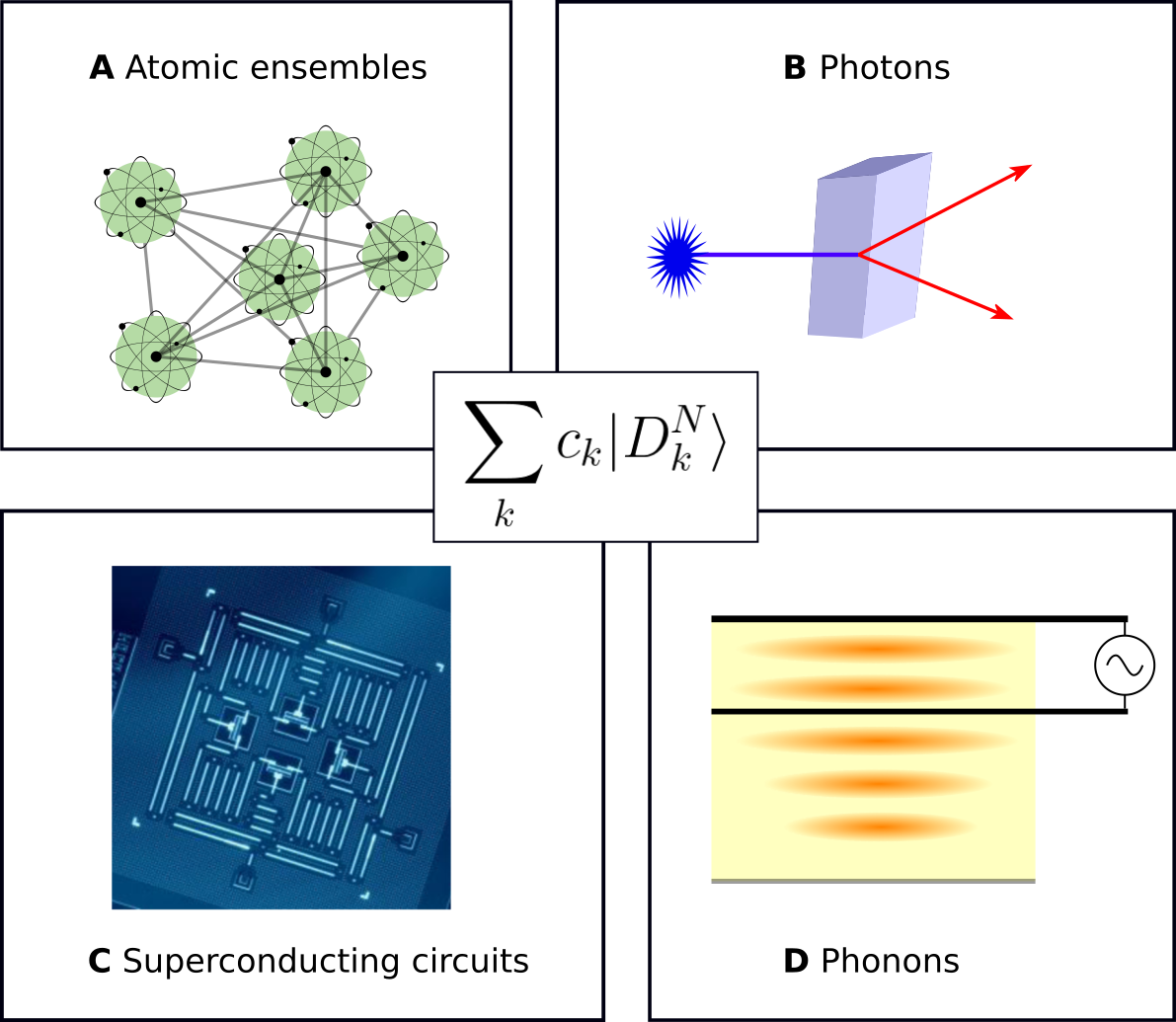}
    \vspace{1mm}
    \caption{Main platforms and techniques used to generate symmetric states in the laboratory. Panel A: atomic ensembles, these include e.g. Bose Einstein condensates. Panel B) illustrates a spontaneous parametric down conversion (SPDC) process -- a standard protocol to generate entangled states of photons. Panel C) displays a transmon qubit in a superconducting device. Panel D) sketches an acoustic resonator hosting phononic modes exited from piezoelectric effects. Source: Panels A,B, original. Panel C~\cite{Gambetta2017}. Panel D inspired from HBAR devices~\cite{Baron13}.}
    \label{fig:platforms} 
\end{figure}

\noindent In this section, we review the most common platforms for realizing symmetric states and highlight recent breakthroughs in their experimental implementation (see also \cite{frerot_probing_2023}). We begin with cold atoms and proceed through trapped ions, photons, superconducting qubits, and mechanical resonators, concluding with quantum algorithms designed for the preparation of symmetric states.
Figure~\ref{fig:platforms} illustrates such platforms, that will be discussed in detail in the following paragraphs. Specifically, in Section~\ref{sec:ColdAtoms} we address platforms involving cold atoms such as cold atomic ensembles (panel A). In turn, photonic platforms (panel B) are discussed in Section~\ref{sec:photons}. Next, in Section~\ref{sec:superconducting} we review the generation of symmetric states in superconducting circuits (panel C). We also present recent experimental advances in the field of mechanical resonators (panel D and Section~\ref{sec:acustic}).








\subsection{Cold atoms}
\label{sec:ColdAtoms}

\subsubsection{Neutral atoms in lattices}

\noindent Ultracold atoms in optical lattices \cite{lewenstein_ultracold_2012} have shown unparallelled potential as quantum simulators \cite{lewenstein2007ultracold, bloch_many-body_2008}, unravelling phenomena such as many-body localization \cite{abanin_colloquium_2019} and scars \cite{bluvstein_controlling_2021}.
In atomic lattices, Rydberg-atom arrays have been used to prepare GHZ states of 20 qubits \cite{omran_generation_2019, song_generation_2019}.
In the context of quantum information processing, adiabatic evolution allows for the preparation of ground states to tackle complex combinatorial optimization tasks \cite{ebadi_quantum_2022, schiffer_adiabatic_2022}. Neutral-atom quantum computers have demonstrated their ability to prepare 6-qubit GHZ states, perform quantum phase estimation and address MaxCut graph problems \cite{graham_multi-qubit_2022}. Lattices up to $280$ qubits have demonstrated their potential even for logical-level control \cite{bluvstein_logical_2024}.

The violation of permutationally-invariant Bell inequalities from one- and two-body correlations \cite{tura2014detecting} in ultracold atoms in a lattice was theoretically investigated in \cite{pelisson_nonlocality_2016}, proposing a scheme that only requires beam-splitting operations, global phase shifts and single-site addressing.


\subsubsection{Atomic ensembles}
\noindent In experimental settings, it is possible to generate thousands of neutral atoms entangled in their spin degrees of freedom. Due to their strong coupling to external fields, atoms are excellent candidates for quantum-enhanced metrological applications \cite{Pezze2018}, where many-body entanglement can be exploited to increase the sensitivity of precision measurements.

In spinor Bose-Einstein condensates (BEC) of ${}^{87}\mathrm{Rb}$ $F=1$ atoms, one can create Dicke states via a parametric process known as ``spin-changing collisions" \cite{lucke_detecting_2014}, where two $m_F=0$ atoms collide producing a $m_F=\pm 1$ pair. In particular, Dicke-like states can be prepared via such processes.
Building on the early theoretical proposal \cite{Sorensen2001}, scalable spin-squeezing for quantum-enhanced magnetometry has been suggested for BECs consisting of more than $10^7$ atoms \cite{muessel_scalable_2014}. In such a way, non-Gaussian entangled states of $380\pm 15$ $^{87}$Rb atoms were generated \cite{strobel_fisher_2014}.

Spin-squeezed states in BECs can be regarded as coherent superpositions of Dicke states with reduced variance compared to a coherent (symmetric product) state \cite{tura2015nonlocality}. This has allowed detection of Bell correlations \cite{tura2014detecting, wagner_bell_2017, fadel_bounding_2017} in mesoscopic systems consisting of approximately $480$ Rb atoms \cite{schmied_bell_2016}. Theoretical works have shown how to certify entanglement \cite{aloy_device-independent_2019, tura_optimization_2019} and Bell correlations depth \cite{baccari_bell_2019} from these experiments.
In \cite{engelsen_bell_2017}, Bell correlations were detected in a thermal ensemble of $5\times 10^5$ $^{87}$Rb atoms at $25 \mu\mathrm{K}$ by preparing an analogous Dicke state superposition via spin squeezing. Higher-order correlators can also be used for detecting non-Gaussian (oversqueezed) quantum states \cite{guo_detecting_2023}.

So far, all these experiments have been carried out in spin-$1/2$ particles. Atomic ensembles of higher spin (particularly, spin-$1$, corresponding to three-level Dicke states) have also been experimentally investigated extensively \cite{stenger_spin_1998, chang_observation_2004, lucke_twin_2011, gross_atomic_2011, Hamley2012}. 
Multilevel collective measurements in spinor BECs can be routinely implemented \cite{ho_spinor_1998, law_quantum_1998,TetsuoOhmi1998} by splitting the population in each atomic level via a magnetic field and estimate the particle number via absorption imaging \cite{chang_observation_2004}. Measurements in other bases can be implemented via unitary transformations prior to measurement \cite{giorda_universal_2003}. Such hyperfine spin states can be accessed via radio-frequency pulses and quadratic Zeeman energy shifts \cite{Hamley2012, sau_spin_2010}.
Entanglement has been detected in such multilevel systems, whereas the experimental detection of Bell correlations in higher-dimensional many-body systems remains open.

Precise atom number detection is a challenge in such platforms. Nevertheless, for mesoscopic atom numbers, techniques have been developed which allow to control the atom number to a single atom level -- for example, up to thousands of atoms in magneto-optical traps \cite{hume_accurate_2013}, or few-particle optically trapped $^6$Li spin $1/2$ fermions \cite{wenz_few_2013,zurn_pairing_2013}.

BEC experiments have also allowed for the preparation of squeezed states in double-well potentials, where Dicke state superpositions can be prepared and entangled across the wells \cite{fadel_spatial_2018, kunkel_spatially_2018, lange_entanglement_2018} (see \cite{fadel_entanglement_2021, fadel_multiparameter_2023} for recent theoretical developments).

\subsubsection{Ultracold atoms in nanostructures}
\noindent Trapping atoms in the vicinity of tapered optical fibres and optical crystals which are band gap materials, offers another possibility for the experimental preparation of many-body symmetric states \cite{gonzalez-tudela_lightmatter_2024, chang_colloquium_2018}. Early theoretical studies noticed the remarkable properties of such systems. In \cite{kien_cooperative_2008}, the possibility of directional guided superradiance from a linear array of distant atoms parallel to a nanofibre was shown. In \cite{zoubi_hybrid_2010}, it was shown that for two parallel lattices an exciton can tunnel from one lattice to the other, creating a quasi-particle called fibre polariton. In \cite{chang_multiatomic_2011} the possibility of a doped resonator array to control the propagation of a photon wave packet and its reflection properties were studied. In \cite{chang_cavity_2012} it was shown that a specially designated impurity atom within the cavity can interact much more strongly with the single photons in the fibre.

All these appealing properties led to proposals for the realization of long-range spin models. In \cite{chang_self-organization_2013}, a rich set of behaviors related to the self-organization of atoms along a nanophotonic waveguide was studied, showing how their properties may be modified simply by changing a few external parameters. This opened the door to use atoms in self-organizing lattices as light-controlled qubits. In \cite{gonzalez-tudela_subwavelength_2015}, atom-atom interactions were engineered in two-dimensional photonic crystals, with interactions that are several orders of magnitude larger than for exchange interactions in free-space lattices, thereby allowing to engineer strong long-range interactions. In \cite{douglas_quantum_2015}, quantum many-body models with cold atoms coupled to photonic crystals were proposed. There, atomic spin degrees of freedom, motion and photons strongly couple over long distances. By eliminating the motional modes, these systems can easily engineer XX-model and transverse Ising models with various interaction properties. Preparation of Dicke states in such platforms has been studied in \cite{franquet_gonzalez_preparation_2013}, showing the possibility to engineer genuine multipartite entangled Dicke states up to $n=10$ atoms.

Experimentally, optical nanofibres demonstrated early on their efficiency as tools to probe atomic fluorescence \cite{nayak_optical_2007}. In \cite{vetsch_optical_2010}, Cs atoms were localized in a 1D optical lattice, about $200 \mathrm{nm}$ above the nanofibre surface, which then was interrogated via a resonant light field sent through the nanofibre. In \cite{goban_demonstration_2012}, an improved method to create optical trap for Cs atoms near the surface of a nanofibre was reported in the context of spectroscopy. In \cite{goban_atomlight_2014}, an integrated optical circuit with a photonic crystal that could both localize and interface atoms using guided photons was reported.

 By means of collective dissipation one can engineer and manipulate atomic and quantum optical states: If a single atom has an enhanced probability to emit a photon into the waveguide, the 1D characteristic of the photonic field makes subsequent atoms have also a large probability to interact with the same photon. This mechanism of emission and reabsorption happens multiple times, leading to collective dissipation. The latter can be used to create quantum many-body states of interest and to trigger multiphoton emission. By placing the atoms in specific equidistant positions in the waveguide, photons couple to a single collective atomic mode, yielding an effective pure Dicke Hamiltonian \cite{dicke_coherence_1954}. Its main effects, superradiance and subradiance \cite{gross_superradiance_1982} become easily accessible experimentally, furthermore ensuring the superradiant emission goes into the guided modes. Beyond the so-called linear response (single-excitation manifold), all symmetric states of $N$ atoms containing $k$ excitations $\ket{D^N_k}$ decay with an enhanced rate of approximately $kN$ for $k \ll N$. In conventional settings, it may be difficult to generate a basis Dicke state $\ket{D^N_k}$ with a fixed $k$ because atomic ensembles with many atoms behave like a linear system. However, atom-waveguide QED characteristics can be exploited to generate $\ket{D^N_k}$ using the following configuration: one single atom is separated several wavelengths apart such that it can then be individually addressed with respect to an ensemble of $N$ atoms that is collectively driven \cite{gonzalez-tudela_deterministic_2015}. Using this mechanism, one can easily map Dicke states superpositions to multiphoton states \cite{porras_collective_2008, gonzalez-tudela_deterministic_2015}. At the expense of making the process probabilistic, the error scaling can be quadratically improved \cite{gonzalez-tudela_efficient_2017} from $O(k/\sqrt{P})$ to $O(k/NP)$, where $P$ is the so-called Purcell factor of the waveguide.



\subsection{Trapped ions}
\noindent Ultracold trapped ions with internal pseudo-spin can interact with each other through phonon excitations. Under some conditions, this allows spin chains to behave with long-range interactions. This allows to entangle ions through their collective quantized motion \cite{cirac_quantum_1995} or through the thermal motion \cite{sorensen_quantum_1999, molmer_multiparticle_1999, sorensen_entanglement_2000}.  Early theoretical proposals suggested the use of inhomogeneous magnetic fields \cite{mintert_ion-trap_2001} 
and appropriately designed laser-ion interactions \cite{porras_effective_2004}
, which were reported in experiments afterwards \cite{friedenauer_simulating_2008, kim_quantum_2010}. The strength of spin interactions is typically inversely proportional to the cube of the distance \cite{porras_effective_2004, deng_effective_2005, hauke_complete_2010, maik_quantum_2012}, but by controlling the phonon dispersion, power laws with exponents between $0.1$ and $3$ can be effectively reached, including 2D arrays of traps \cite{britton_engineered_2012}. Such controllable interactions have also been reported in quantum integer-spin chains \cite{senko_realization_2015}. Trapped ion chains with long range SU(3) interactions have a deep relationship with quantum chaos \cite{gras_quantum_2013}, which has also been studied in the context of Bell correlations \cite{aloy_bell_2024} (see also \cite{aloy_deriving_2024, MullerRigat2024, gras_trapped-ion_2014}).

Most experiments have employed one of two types of trapping mechanisms for charged particles: Paul traps, which uses modulated electric fields and Penning traps, which combine static electric and magnetic fields. Paul traps typically confine tens of ions in a 1D geometrical configuration, resulting from the interplay between the trapping potential and Coulomb repulsion. In Penning traps, hundreds of atoms form a 2D array, but the static magnetic field makes them rotate at a given frequency. These reasons imply that single-particle addressability is much easier in Paul traps through laser pulses, whereas Penning traps are much more suited to collective operations, as in atomic ensembles. Compared to the latter, ion traps are more limited in the number of particles they can handle, although they do not suffer from particle losses as in atomic ensembles. Furthermore, the more ions in the system, the more difficult individual addressability becomes, as motional eigenmodes cluster in frequency, decreasing the resulting fidelity.

In the context of symmetric state preparation, there exist theoretical proposals for the deterministic generation of arbitrary symmetric states in ion chains \cite{hume_preparation_2009, lamata_deterministic_2013, ivanov_creation_2013}.
GHZ states up to $6$ \cite{leibfried_creation_2005} and  $14$ qubits have been reported experimentally in \cite{monz_14-qubit_2011}. In \cite{haffner_scalable_2005}, W states of up to $8$ ions have been prepared and detected, both via full tomographic quantum state reconstruction and via generalized spin-squeezing inequalities \cite{korbicz_generalized_2006}. In the context of device-independent genuine multipartite entanglement detection \cite{aloy_device-independent_2019, tura_optimization_2019}, as well as multipartite nonlocality \cite{baccari_bell_2019}, both phenomena were observed for a GHZ of six entangled ions \cite{barreiro_demonstration_2013}. In \cite{friis_observation_2018}, a fully-controllable $20$-qubit register showed genuine multipartite entanglement depth up to $5$ particles, by studying out-of-equilibrium dynamics of an Ising-type XY Hamiltonian, which was further quantified via R\'enyi entropies \cite{brydges_probing_2019}. Bell nonlocality has further been demonstrated in systems between three and seven ions \cite{lanyon_experimental_2014}, in the context of measurement-based quantum computation \cite{briegel_measurement-based_2009} and through violation of the Mermin-Klyshko Bell inequalities \cite{mermin_extreme_1990, belinskii_interference_1993, scarani_spectral_2001}.

Effective long-range Ising models with and without external transverse fields have been prepared in trapped ion systems. An iconic model in cavity QED is the Dicke model \cite{dicke_coherence_1954}, which describes the coupling between a large spin system and an oscillator. In suitable parameter regimes, the phonons can be adiabatically eliminated, leading to an effective spin Lipkin model. This has been experimentally implemented in a self-assembled two-dimensional Coulomb crystal in a Penning trap of about $70$ ${}^9\mathrm{Be}^+$ ions in the presence of an external transverse field \cite{safavi-naini_verification_2018}. Recently, simple protocols involving solely global rotations and globally-applied non-linear phase gates, without the need of individual addressing or ancillary qubits, have been proposed in \cite{bond_efficient_2023}. Such schemes estimate fidelities above $95\%$ for typical experimental noise levels (see also \cite{gutman_universal_2024}). In \cite{britton_engineered_2012, bohnet_quantum_2016}, spin-squeezed and over-squeezed states were engineered through long-range homogeneous Ising interactions. In \cite{gilmore_quantum-enhanced_2021}, spin-motion entangled states were prepared using spin-dependent optical dipole forces with around $150$ trapped ions.

\subsection{Photons}
\label{sec:photons}

\noindent Entangled states of photons can be prepared in the lab via the interaction of light with non-linear crystals. This has led to many experiments involving few photons \cite{pan_multiphoton_2012}.

Photons can be entangled in their polarisation degree of freedom. 
A method for generating all symmetric Dicke states, either in the long-lived internal levels of $n$ massive particles, or in the polarization degrees of freedom using only linear optics was proposed in \cite{thiel_generation_2007}. Experimental entanglement for graph states of six qubits has been demonstrated \cite{lu_experimental_2007}. Processes like spontaneous parametric down-conversion split a high-energy pump photon into two lower-energy photons thus producing an entangled photon pair \cite{lamas-linares_stimulated_2001}. Due to the probabilistic nature of SPDC processes, the preparation of multipartite entangled states faces serious scalability challenges. Nevertheless, GHZ states of 3 \cite{bouwmeester_observation_1999, pan_experimental_2000}, 8 \cite{huang_experimental_2011, yao_observation_2012}, 10 \cite{wang_experimental_2016, chen_observation_2017} and 12 photons have been recently prepared \cite{zhong_12-photon_2018}.
 Higher-order SPDC processes can also be harnessed to directly prepare tripartite entangled states \cite{chang_observation_2020}.

Entanglement of many photons via SPDC can be certified via the PPT criterion up to $12$ photons and with other criteria up to hundreds of photons \cite{eisenberg_quantum_2004}. Bound entanglement of photons has been detected \cite{hiesmayr_complementarity_2013} for the two-qutrit Horodecki state \cite{horodecki_mixed-state_1998}. 
In \cite{marconi2024entanglement}, a protocol to generate and certify the entanglement of two-qudit DS states was presented, leveraging on the concept of path entanglement between photons.

Experimental preparation of Dicke states has been achieved in the lab for four \cite{kiesel_experimental_2007, hiesmayr_observation_2016} and six \cite{wieczorek_experimental_2009, prevedel_experimental_2009} photons. More recently, three-photon polarization-entangled W state with the assistance of cross-Kerr nonlinearities and classical feedforward was generated \cite{dong_nearly_2016} (see also \cite{sun_schrodinger_2019}).

Using a so-called Rydberg superatom, consisting of an atomic ensemble of Rydberg atoms under Rydberg blockade \cite{saffman_quantum_2010}, one can bypass the probabilistic hindrance of SPDC and generate up to six qubits multipartite entanglement \cite{yang_sequential_2022}, as well as GHZ states up to $14$ photons or linear cluster states up to $12$ photons with fidelities lower-bounded by $76\%$ and $56\%$, respectively \cite{thomas_efficient_2022}.

\subsubsection{Integrated photonics}
\noindent Integrated quantum photonics are based on the ``lab-on-a-chip" approach \cite{adcock_advances_2021, wang_integrated_2020}. Their steady increase in scalability over the last years has allowed progress from few-component circuitry, occupying centimetre-scale footprints, capable only to generate, process, and detect of quantum states of just two photons, to fully-programmable devices on the order of thousands of optical components, at the millimetre-scale footprint, with integrated generation of multiphoton states \cite{taballione_20-mode_2023, bao_very-large-scale_2023}.

In \cite{matthews_manipulation_2009}, four-photon entangled states were prepared and in \cite{llewellyn_chip--chip_2020}, four-photon GHZ states were produced. Four-photon graph states were produced in a silicon chip in \cite{adcock_programmable_2019}. By means of integrated frequency combs, four-partite entanglement in time-bins had also been produced via the product of biphoton Bell states \cite{reimer_generation_2016}. In \cite{grafe_-chip_2014}, single-photon $\ket{W}$ states up to $16$ spatial modes were reported.

Exploiting both the polarization and momentum degrees of freedom, GHZ (N00N) states of $6$, $8$ and $10$ photons were prepared in \cite{gao_experimental_2010}. In \cite{wang_18-qubit_2018}, $18$-qubit GHZ states were prepared by means of combining three degrees of freedom of six photons (path, polarization and orbital angular momentum).

Recent advances in the generation of multidimensional entanglement have also been demonstrated. In \cite{malik_multi-photon_2016}, a multipartite entangled GHZ-like state was prepared in ${\cal H} = \mathbb{C}^3 \otimes \mathbb{C}^3 \otimes \mathbb{C}^2$. In \cite{wang_multidimensional_2018}, multidimensional quantum entanglement with large-scale integrated optics was demonstrated for bipartite states of dimensions up to $15\times 15$.


\subsubsection{Continuous-variable entanglement}
\noindent Multipartite entangled states in continuous-variable systems have been studied through $\hat{x}$ and $\hat{p}$ quadratures \cite{braunstein_quantum_2005}. Deterministic generation of continuous-variable cluster states  containing more than $10^4$ time-domain modes \cite{yokoyama_ultra-large-scale_2013} and up to $60$ frequency-domain modes \cite{gerke_full_2015, chen_experimental_2014, roslund_wavelength-multiplexed_2014, cai_multimode_2017} were investigated leveraging on frequency combs.
By removing a single photon in a mode-selective manner from a Gaussian state \cite{weedbrook_gaussian_2012}, three-mode non-Gaussian states can be prepared \cite{ra_non-gaussian_2020}. Beyond photon-subtraction protocols, photon-addition has also been used to generate optical cat states \cite{chen_generation_2023}.

\subsection{Superconducting qubits}
\label{sec:superconducting}

\noindent Superconducting qubits offer a versatile platform for the preparation of symmetric states. Deterministic generation of GHZ state of 18 qubits was achieved in \cite{song_generation_2019}. In \cite{wang_probing_2025} Bell-operator correlations were probed in a chip of $73$ qubits, certifying genuine multipartite Bell-operator correlations of depth up to $24$. To this end, nonlocality depth witnesses based on permutationally-invariant Bell inequalities were used \cite{baccari_bell_2019} (see also \cite{aloy_device-independent_2019, tura_optimization_2019} for their device-independent entanglement witness analogue). The single-particle addressability allowed to certify a degree of correlations stronger than those of experiments relying solely on collective observable measurements. Although it is possible to break the permutation symmetry in these methods \cite{li_improved_2024}, e.g. via quantum neural states \cite{deng_machine_2018} or restricting the symmetry to just translational invariance
\cite{tura_translationally_2014, tura_energy_2017, emonts_effects_2024, hu_tropical_2022, hu_characterizing_2024}, symmetric states still offer a valuable variational ansatz \cite{Aloy2021}.

In the context of scaling up superconducting circuits, quantum local area networks offer promising possibilities.
Cryogenic links have allowed for fast and high-fidelity measurements such that even a loophole-free Bell test was possible across a $30$m distance \cite{storz_loophole-free_2023}.
Such a proof-of-concept milestone brings closer to realistic protocols such as cross-platform verification \cite{knorzer2023cross}, where controlled-permutation tests are carried over different devices \cite{koczor_exponential_2021}. 
Protocols for simultaneous eigenstate preparation \cite{schiffer_quantum_2025} and energy filtering \cite{liu_preparing_2025} have also been investigated, where instead of permuting the system registers, controlled permutations are applied on ancillary qubits.
Multipartite entanglement measures, such as concentratable entanglement \cite{beckey_computable_2021, beckey_multipartite_2023, coffman_local_2024} has also been generalized \cite{liu_generalized_2024} to multiple devices via such controlled permutation tests, showing e.g. that three devices lead to probabilistic entanglement concentration into W states, and conjecturing new multipartite Tsallis-entropic inequalities.



Preparation of Dicke states has also been investigated in the context of spin ensembles using quantum phase estimation. That offers a non-deterministic protocol by collectively coupling the qubits to an ancillary qubit via $\sigma_z \otimes \sigma_z$ interaction, using $\lceil \log N \rceil +1$ ancillary qubit measurements, a scheme particularly suited for NV-centers coupled to a single superconducting flux qubit \cite{wang_preparing_2021}.

Hybrid quantum circuits combining superconducting circuits and other quantum systems could harness the advantages and strengths of both platforms \cite{xiang_hybrid_2013}. For donor nuclear spins in silicon, theoretical proposals have shown that Dicke states of $10$-$20$ qubits could be prepared with realistic parameters with high fidelity 
\cite{luo_deterministic_2012}. Silicon hybrid quantum processors with robust long-distance qubit couplings could be possible experimental implementations
\cite{tosi_silicon_2017}. These are based on the Kane proposal for a nuclear-spin quantum computer in silicon, using short-range exchange interactions between donor-bound electrons to mediate an effective inter-nuclear coupling of $\sim 100$kHz at $\sim15$ nm distance \cite{kane_silicon-based_1998}. The preparation of spin eigenstates in spintronic quantum computing architectures, including Dicke states, has been discussed in \cite{sharma_preparation_2021}.

\subsection{Mechanical resonators in circuit quantum acoustodynamics}
\label{sec:acustic}
\noindent Mechanical resonators confine many high-quality-factor bosonic modes into a small volume, allowing for an easy integration with different quantum systems\cite{chu_perspective_2020}. Via a beamsplitter interaction between two acoustic modes, families of maximally entangled qubit states, incuding $\ket{D^2_1}$ can be created \cite{von_lupke_engineering_2024}.

By means of phonon transmission, surface acoustic waves serve as quantum transducers that can propagate quantum information encoded in various quantum degrees of freedom \cite{schuetz_universal_2015, delsing_2019_2019}. This phonon-mediated quantum state transfer allows for remote qubit entanglement \cite{bienfait_phonon-mediated_2019} and for the creation and control of multi-phonon Fock states in bulk acoustic-wave resonators \cite{chu_creation_2018,PhysRevLett.134.180801}.

Strong coupling betweeen surface-acoustic wave resonators and superconducting qubits can be engineered to use the qubit to control and measure quantum states in the mechanical resonator \cite{satzinger_quantum_2018}.

Theoretical proposals showcase the potential of solid-state platforms based on surface acoustic waves for trapping, cooling and controlling charged particles \cite{schuetz_acoustic_2017}. Electron lattices  can serve as a platform for the simulation of quantum many-body systems, with lattice parameters that can be tuned in real time \cite{knorzer_solid-state_2018}. Recent proposals have considered the simulation the Dicke lattice model with an array of coupled mechanical resonators that homogeneously interact with a group of trapped BECs \cite{leng_simulating_2022}.

\subsection{Quantum algorithms for Dicke state preparation}
\begin{figure}[h]
\centering
\includegraphics[width=1.\linewidth]{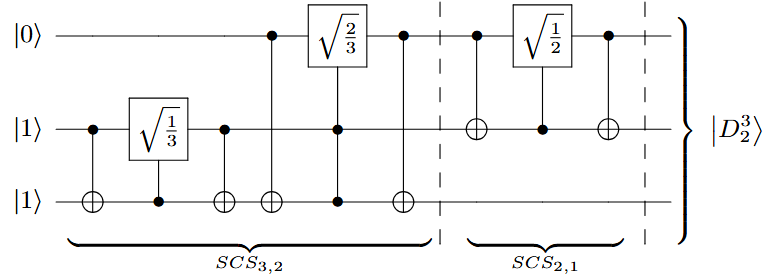}
\caption{A deterministic quantum algorithm to prepare the Dicke state $\ket{D_2^3}$ of $N=2$ qubits with $k=3$ excitations. The number of initial qubits in the state $\ket{1}$ determines the number of excitations in the Dicke state. The $\sqrt{\frac{a}{b}}$ gates represent Y-rotations $R_y(2 \cos^{-1} \sqrt{\frac{a}{b}})$. Source: inspired from \cite{bartschi_deterministic_2019}}
    \label{fig:deterministicalgorithmdicke}
\end{figure}

\noindent Dicke states can also be prepared in quantum computers via specifically designed quantum circuits of different complexity. There exist quantum algorithms that probabilistically prepare Dicke states in time $O(N\log N)$ with $O(\log N)$ ancillary qubits \cite{childs_finding_2002}.
Deterministic algorithms for the preparation of qubit Dicke states 
$\ket{D^N_k}$ with complexity $O(kN)$ in the number of gates and $O(N)$ in quantum circuit depth, without ancillary qubits have been proposed in \cite{bartschi_deterministic_2019} such as the one that can be seen in Figure \ref{fig:deterministicalgorithmdicke}. These algorithms can prepare symmetric states in $O(N^2)$ gates and $O(N)$ circuit depth. Follow-up works have investigated the preparation of Dicke states in cloud quantum computers \cite{mukherjee_actual_2020}, as well as divide-and-conquer approaches \cite{aktar_divide-and-conquer_2022} and the effect of hardware connectivity limitations \cite{bartschi_short-depth_2022} ($O(k\log (N/k))$ for all-to-all connectivity vs. $O(\sqrt{kN})$ in a grid), which is an improvement over earlier proposals \cite{kaye_quantum_2001}. Recently, a new deterministic quantum circuit for the generation of Dicke states was proposed, whose depth scales as ($O(\log(k)\log (N/k)+k)$ and which does not require the use of ancillary qubits \cite{yuan2025depthefficientquantumcircuitsynthesis}.
For specific classes, such as W states, the complexity can be reduced to $O(\log N)$ circuit depth and $O(N)$ entangling (CNOT) gates \cite{cruz_efficient_2019}.
Similarly, dedicated quantum circuits for the preparation of qudit Dicke states have been recently proposed \cite{nepomechie_qudit_2023, nepomechie_spin-s_2024, raveh_q-analog_2024}, which serve as preliminary steps for more general classes of quantum states related to integrable models, such as Bethe ans\"atze \cite{raveh_deterministic_2024, li_bethe_2022, dyke_preparing_2022, sopena_algebraic_2022, ruiz_bethe_2023,ruiz2025betheansatzquantumcircuits}.

The quantum Schur transform maps between the computational basis of $(\mathbb{C}^d)^{\otimes N}$ and the so-called Schur basis, which simultaneously diagonalizes the commuting actions of both the symmetric group of $N$ elements ${\mathcal G}_N$ and the unitary group of order $d$, ${\mathcal U}_d$. Dicke states are perhaps the most notable subset of the Schur basis. Polynomial-time quantum circuits for the Schur transform were proposed in \cite{harrow_applications_2005, bacon_quantum_2007}, with recent extensions considered in \cite{wills_generalised_2024} and \cite{nguyen_mixed_2023}. The  Schur transform can be seen as a method of compression of quantum states and therefore, uncompression can be used for state preparation. This more general change of basis allows for a circuit depth of $O(N \mathrm{polylog}(N/\varepsilon))$ with the same order of entangling gates and $O(\log(N/\varepsilon))$ auxiliary qubits \cite{bacon_efficient_2006}. That can be achieved at a cost of $O(N^2)$ without auxiliary qubits for preparing symmetric states \cite{plesch_efficient_2010}.

It is worth noting that general-purpose algorithms for sparse quantum state preparation can be used for $d$-sparse Dicke states, which for qubits amounts to a $d={N \choose k}$ sparsity. Using \cite{zhang_quantum_2022}, that would yield a circuit depth of $O(\log(dN))$, with $O(d N \log(dN))$ CNOT gates and $O(dN\log d)$ auxiliary qubits \cite{bartschi_short-depth_2022}. 

Counterdiabatic driving has been explored for spin squeezing Dicke state preparation \cite{opatrny_counterdiabatic_2016} by turning the initial ground state of a linear Hamiltonian gradually into the ground state of a quadratic Hamiltonian, which corresponds to the selected Dicke state. Compensating operators can be engineered to supress diabatic transitions that would otherwise occur if the changes were not slow. Such shortcuts to adiabaticity in the Lipkin-Meshkov-Glick Hamiltonian were studied in \cite{takahashi_transitionless_2013, campbell_shortcut_2015}.


Very recently, protocols to prepare quantum many-body states with quantum circuits assisted by LOCC have been proposed \cite{piroli_approximating_2024}. If one drops the requirement of exact preparation, substantial saves in computational resources can be achieved. In particular, $N$-qubit Dicke states with $k$ excitations can be approximately prepared with infidelity $\varepsilon$ with $O(1)$ circuit depth, $O(l_{k,\varepsilon})$ ancillary qubits and $O(\sqrt{k})$ repetitions, where $l_{k,\varepsilon} = \max\{\log_2(4k), 1+\log_2\ln (\sqrt{8\pi k}/\varepsilon)\}$. One can easily trade-off depth with number of ancillas, e.g. obtaining $O(l_{k,\varepsilon})$ circuit depth and $O(1)$ ancillas for the same number of repetitions. If one wants to keep the number of repetitions exactly $1$, then it is possible to achieve a circuit depth of $O(k^{1/4}l^2_{k,\varepsilon})$ and $1+l_{k,\varepsilon}/N$ ancillas. Such algorithms become especially relevant in present-day quantum devices, where quantum circuits can be assisted by mid-circuit measurements and feed-forward operations \cite{sukeno_quantum_2024,baumer_efficient_2024, chen_nishimori_2025, iqbal_non-abelian_2024}.

\subsection{Dicke states in physical systems}
\noindent Dicke states appear naturally as eigenstates of some physical systems. Most notably, in the so-called Lipkin-Meshkov-Glick Hamiltonian \cite{LMG1,LMG2,LMG3},
$$H = -\frac{\lambda}{n}\sum_{i<j}(X_iX_j + \gamma Y_iY_j) - h \sum_i Z_i,$$
as one tunes the external magnetic field $h$, in the isotropic phase $\gamma = 1$, $\ket{D^N_k}$ appear as ground states of the LMG model for qubits. The LMG Hamiltonian is one of the few one-dimensional long-range quantum systems that has been exactly solved, including its entanglement entropy \cite{latorre_entanglement_2005}, and it was originally introduced in the context of nuclear physics, to describe a system of $n$ fermions in two levels, each $n$ times degenerate. Moreover, there has been considerable interest in studying several entanglement measures of the ground states of this model \cite{Or_s_2008,Wichterich_2010}. Similar investigations have been carried out for generalized isotropic LMG models \cite{carrasco_generalized_2016}, where qudit Dicke states appear as its ground state for suitable parameter choices.


It is possible to approximately prepare Dicke state superpositions in the form of tensor networks \cite{cirac_matrix_2021}, in particular, matrix product states \cite{Aloy2021}. The Parent Hamiltonian construction allows for the construction of a local, frustration-free, Hamiltonian whose ground state is the one described by the tensors, as long as certain technical conditions such as injectivity are met. The choice of such Parent Hamiltonian is not unique and recent works have shown how to optimize for its locality \cite{giudici_locality_2022} or spectral gap \cite{rai_spectral_2024, rai_hierarchy_2024, chen_boosting_2024}.








\section{Concluding remarks}
\label{sec:Conluding Remarks}

\noindent Symmetric quantum states play a central role in the landscape of quantum information science. 
Their intrinsic mathematical properties as well as the versatility of their applications have generated significant theoretical and experimental progress over the past two decades. In this review, we have highlighted their fundamental features, presented fundamental techniques for their characterisation and verification, and surveyed the range of tasks in which they provide distinct advantages.

Yet, despite this progress, several important open questions remain, pointing to promising avenues for future research. One such challenge is the general characterization of PPT-entangled symmetric states, both in bipartite and multipartite systems. In the case of bipartite diagonal symmetric (DS) states, it has recently been shown that separability is equivalent to a cyclic variation of the complete positivity problem \cite{gulati2025entanglement}, and PPT-entangled DS states have been fully characterized for dimension $d \leq 5$. For multipartite diagonal symmetric states much less is known. While for N=3 and d=3 there exist PPT-Bound DS states  \cite{PrivateCommunicationNechita}, the complexity of this task becomes significantly harder as the number of parties or local Hilbert dimension increases \cite{PrivateCommunicationJordi}, indicating the need for new mathematical tools to address this problem.
Studying the properties of  non-decomposable entanglement witnesses for symmetric states \cite{marconi2021entangled,champagne2022spectral} is also of interest. Aside from their connection with the cone of copositive matrices, advances in this area will led to the discovery of novel properties of positive, but not completely positive, maps. 
Symmetric states may also play a key role in addressing the conjecture, originally formulated in \cite{sanpera2001schmidt}, that no PPT state in $\mathbb{C}^{d} \otimes \mathbb{C}^{d}$ can have maximal Schmidt number (that is, equal to $d$). While the conjecture has been proven for $d=3$ \cite{yang2016all}, extending this result to higher dimension remains a challenging task. A closely related topic is the use of symmetric states to prove the
PPT$^2$ conjecture \cite{christandl2012ppt, christandl2019composed}-- stating  that the composition of two PPT channels is entanglement breaking.
For bipartite systems of qudits, this conjecture holds trivially for $d=2$, as a consequence of the PPT criterion. Moreover, in \cite{chen2019positive,christandl2019composed}, it has been proven for $d=3$, but remains open for $d>3$. When restricting to symmetric states, it was shown in \cite{singh2022ppt} that the conjecture holds true for Choi-type maps, which include bipartite DS states as a special case. However, an analogous result for symmetric, but not DS, states has yet to be established.

Another fundamental question concerns the identification of maximally entangled multipartite pure states within the symmetric subspace. Since there is no unique definition of maximal entanglement in the multipartite setting, one must first select a suitable entanglement measure. Promising results in this direction have been reported in \cite{aulbach2010maximally, martin2010multiqubit}, which employ the geometric measure of entanglement. These works also suggest a deeper connection between highly entangled symmetric states and their Majorana representation, a topic which deserves further analysis.

On a more applied level, while symmetric states have proven to be useful tools in areas such as quantum metrology and error correction, their potential in quantum communication tasks remains relatively unexplored, with the sole exception of GHZ and W states. Investigating their utility in protocols such as quantum key distribution, teleportation, or dense coding -- particularly under practical constraints like limited control or imperfect symmetry -- could uncover new roles for symmetric states in distributed quantum technologies.

In summary, while symmetric quantum states are already indispensable in a variety of quantum information tasks, their full potential is far from being realized. Addressing the open questions outlined above is likely to yield both foundational insights and practical advancements, further consolidating the role of symmetric states in quantum technologies. 

To conclude, we would like to acknowledge that, in spite of our best efforts to maintain an up-to-date list of references, certain works may have been inadvertently omitted, and for this, we offer our sincere apologies.

\noindent\textbf{Acknowledgements}
\newline
Carlo Marconi acknowledges support from the European Union - NextGeneration EU, "Integrated infrastructure initiative in Photonic and Quantum Sciences" - I-PHOQS [IR0000016, ID D2B8D520, CUP B53PGC22001750006]
\newline
Guillem Müller-Rigat acknowledges support from: European Research Council AdG NOQIA; MCIN/AEI (PGC2018-0910.13039/501100011033, CEX2019-000910-S/10.13039/501100011033, Plan National FIDEUA PID2019-106901GB-I00, Plan National STAMEENA PID2022-139099NB, I00, project funded by MCIN/AEI/10.13039/501100011033 and by the “European Union NextGenerationEU/PRTR" (PRTR-C17.I1), FPI); QUANTERA DYNAMITE PCI2022-132919, QuantERA II Programme co-funded by European Union’s Horizon 2020 program under Grant Agreement No 101017733; Ministry for Digital Transformation and of Civil Service of the Spanish Government through the QUANTUM ENIA project call - Quantum Spain project, and by the European Union through the Recovery, Transformation and Resilience Plan - NextGenerationEU within the framework of the Digital Spain 2026 Agenda; Fundació Cellex; Fundació Mir-Puig; Generalitat de Catalunya (European Social Fund FEDER and CERCA program; Barcelona Supercomputing Center MareNostrum (FI-2023-3-0024); Funded by the European Union (HORIZON-CL4-2022-QUANTUM-02-SGA PASQuanS2.1, 101113690, EU Horizon 2020 FET-OPEN OPTOlogic, Grant No 899794, QU-ATTO, 101168628), EU Horizon Europe Program (This project has received funding from the European Union’s Horizon Europe research and innovation program under grant agreement No 101080086 NeQSTGrant Agreement 101080086 — NeQST); ICFO Internal “QuantumGaudi” project; European Union - NextGeneration EU (PRTR-C17,l1), PPCC-MCIN-L5.
\newline
 Jordi Romero-Pallejà acknowledges financial support from Ministerio de Ciencia e Innovación of the Spanish Goverment FPU22/01511 and from the Spanish MICIN project PID2022-141283NB-I00.
\newline
Jordi Tura acknowledge the support received
from the Dutch National Growth Fund (NGF), as part of the Quantum Delta NL programme as well as support received from the European Union’s Horizon
Europe research and innovation programme through the
ERC StG FINE-TEA-SQUAD (Grant No. 101040729).
\newline
 Anna Sanpera acknowledges financial support from Ministerio de Ciencia e Innovación of the Spanish Goverment with funding from European Union NextGenerationEU (PRTR-C17.I1) and by Generalitat de Catalunya and from the European Commission QuantERA grant ExTRaQT (Spanish MICIN project PCI2022-132965), by the Spanish MICIN (project PID2022-141283NB-I00) with the support of FEDER funds, and by the Ministry for Digital Transformation and of Civil Service of the Spanish Government through the QUANTUM ENIA project call - Quantum Spain project, and by the European Union through the Recovery, Transformation and Resilience Plan - NextGeneration EU within the framework of the Digital Spain 2026 Agenda.
 The views and opinions expressed here are solely
those of the authors and do not necessarily reflect those of the funding institutions. Neither of the funding institution can be held responsible for them.

\bibliographystyle{apsrev4-2}
\bibliography{Bibliography/lib,Bibliography/lib_carlo,Bibliography/lib_guillem,Bibliography/references2}

\end{document}